\PassOptionsToPackage{dvipsnames}{xcolor}
\documentclass[manuscript,screen]{acmart}

\usepackage[vlined]{algorithm2e}
\usepackage{algpseudocode}
\usepackage{bbm}
\usepackage{mathtools}
\usepackage{mleftright}
\usepackage{subfig}

\usepackage{tikz}
\usetikzlibrary{arrows.meta}

\usepackage{pgfplots}
\usepackage{pgfplotstable}
\pgfplotsset{compat=1.9}

\usepackage{amsthm}
\newtheorem{remark}{Remark}[section]

\newcommand{\bzero}{{\boldsymbol 0}}

\newcommand{\bb}{{\boldsymbol b}}
\newcommand{\bc}{{\boldsymbol c}}
\newcommand{\bm}{{\boldsymbol m}}
\newcommand{\bn}{{\boldsymbol n}}
\newcommand{\bv}{{\boldsymbol v}}
\newcommand{\bx}{{\boldsymbol x}}
\newcommand{\bu}{{\boldsymbol u}}
\newcommand{\bP}{{\boldsymbol P}}
\newcommand{\bR}{{\boldsymbol R}}
\newcommand{\bU}{{\boldsymbol U}}

\newcommand{\polf}{{\mathbbm f}}
\newcommand{\polI}{{\mathbb I}}

\newcommand{\tr}{^{\mathsf T}}

\newcommand\DIV{\nabla\cdot\,}

\newcommand\GRAD{\nabla}

\newcommand\CROSS{\times}
\newcommand\SCAL{\cdot}

\newcommand\diff{{\mathrm{d}}}

\newcommand\Domain{\Omega}
\newcommand\R{{\mathbb R}}

\newcommand\TT{\mathcal{T}}
\newcommand\NN{\mathcal{N}}
\newcommand\Qo{\mathbb{Q}_{1}}

\newcommand\Ii{\mathcal{I}(i)}
\newcommand\bUni{\bU^n_i}
\newcommand\bUnj{\bU^n_j}
\newcommand\bUnij{\overline{\bU}^n_{ij}}

\begin{document}

\title[Efficient parallel second-order Euler solver]%
  {Efficient parallel 3D computation of the
  compressible Euler equations with an invariant-domain preserving
  second-order finite-element scheme}

\author{Matthias Maier}
\email{maier@math.tamu.edu}
\orcid{0000-0002-4960-5217}
\affiliation{%
  \institution{Department of Mathematics, Texas A\&M University}
  \streetaddress{3368 Blocker Building}
  \city{College Station}
  \state{TX}
  \postcode{77843}
  \country{USA}
}

\author{Martin Kronbichler}
\email{kronbichler@lnm.mw.tum.de}
\affiliation{%
  \institution{Institute for Computational Mechanics, Technical University of Munich}
  \streetaddress{Boltzmannstr.~15}
  \postcode{85748}
  \city{Garching}
  \country{Germany}
}

\begin{abstract}
  We discuss the efficient implementation of a high-performance second-order
  collocation-type finite-element scheme for solving the compressible Euler
  equations of gas dynamics on unstructured meshes. The solver is based on the \emph{convex
    limiting} technique introduced by Guermond et
  al.~(SIAM J. Sci. Comput. 40, A3211--A3239, 2018). As such it is \emph{invariant-domain
  preserving}, i.\,e., the solver maintains important physical invariants
  and is guaranteed to be stable without the use of ad-hoc tuning parameters.
  This stability comes at the expense of a significantly more involved
  algorithmic structure that renders conventional high-performance
  discretizations
  challenging.
  We develop an algorithmic design that allows SIMD vectorization of the
  compute kernel, identify the main ingredients for a good node-level
  performance,
  and report excellent weak and strong scaling of a hybrid
  thread/MPI parallelization.
\end{abstract}



\begin{teaserfigure}
  \centering
  \includegraphics[width=0.7\textwidth]{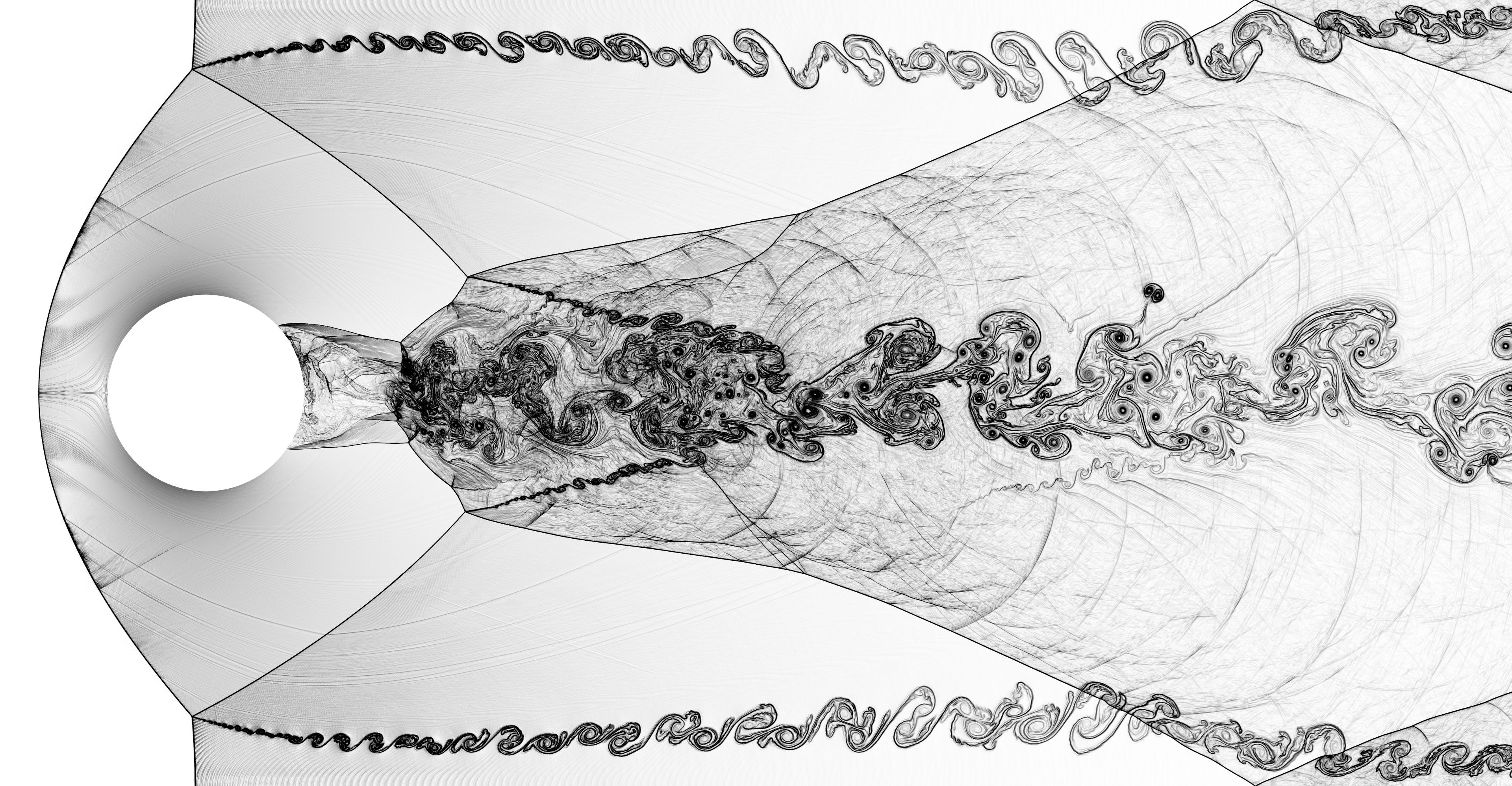}
  \caption{2D Mach 3 compressible Euler flow around a disc; 38\,M
    unstructured $\mathcal Q_1$ nodes, Schlieren-like plot at $t=3.5$.}
  \Description{figure description}
  \label{fig:cyl}
\end{teaserfigure}

\maketitle


\section{Introduction}
\label{sec:introduction}

The appropriate discretization and simulation of the compressible Euler
equations of gas dynamics is an ongoing and intensely discussed
debate~\cite{Zalesak_2005,Kuzmin_2005,Barlow_2016,gnpt_second_order_2018}.
This is in contrast to, for example, the incompressible Navier-Stokes
equations for which a much more complete mathematical solution theory is
available that establishes a common
framework to assess the quality and approximation property of fluid
solvers at least for the pre-turbulent regime~\cite{Schafer_1996}.
The lack of an
accepted solution theory for the Euler equations allows for
considerable freedom in the the notion what constitutes a good
computational approximation (see for
example~\cite{Zalesak_2005,Kuzmin_2005}) and thus in the choice of
discretization scheme.
Consequently, discretization schemes that allow for a high arithmetic
intensity and good parallel scaling have received a high level of attention
during the last decade. An important example are high-order discontinuous
Galerkin (DG) discretizations~\cite{Wang_2013, Witherden_2014,
Kronbichler_2019} with
some form of flux reconstruction and appropriate flux/slope
limiters~\cite{Cockburn_1989}. However, in the transonic and supersonic
regime found in certain shock-hydrodynamics applications, the use of
variational schemes might become questionable due to the lack of pointwise
stability properties---at least without the perpetual hunt for the right
shock capturing technique~\cite{Barlow_2016}.

In this publication we want to entertain a different approach. Instead of
starting with a high-order discretization and
then constructing ad-hoc limiting techniques for solving certain benchmark
problems, we instead start with the mathematical description of a
second-order collocation-type finite-element scheme that is based on the
convex-limiting technique pioneered by Guermond et
al.~\cite{gp_first_order_2016, gp_fast_riemann_2016, gp_second_order_2018,
gnpt_second_order_2018}. The methodology is invariant-domain preserving
\cite{gp_second_order_2018}. This means that in addition to the usual
notion of hyperbolic conservation (regarding density, momentum and total
energy), a number of important physical invariance principles are
maintained strongly: positivity of the density and internal energy and a
local minimum principle on the specific entropy (see
Section~\ref{subsec:specific-entropy-limiter}). The method is guaranteed to
be stable without the use of any ad-hoc tuning parameters. This stability
comes at the expense of a significantly more involved algorithmic
structure. Taking the mathematical properties of the convex-limited
collocation-type continuous Galerkin scheme as a given, the contribution of
the present work is the
identification of data structures and algorithms that
make it run fast on modern hardware, and characterize the proposed
computing kernels in an academic setting. In detail, our
contributions with the current work can be summarized as follows:
\begin{itemize}
  \item
    We describe the algorithmic structure of a  second-order
    collocation-type finite-element scheme for solving the compressible
    Euler equations of gas dynamics. Our solver is based on a slight
    modification of~\cite{gnpt_second_order_2018} suitable for SIMD
    vectorization to render it highly process and thread parallelizable. A
    high degree of instruction-level vectorization can be achieved for
    a nonlinear convex-limiting scheme that involves a large number of
    root-finding problems with transcendental functions as building blocks.
    Our approach is based on explicit vectorization using the C++ template
    mechanism and operator overloading as a high-level user
    interface~\cite{dealiicanonical,Kronbichler_2012}, as well as on
    algorithmic design that avoids branching on data.
  \item
    We comment on optimization strategies to achieve excellent scaling
    characteristics and absolute performance, such as, avoiding index
    translations, cache-optimized traversal of data structures, using
    point-to-point MPI communication, and efficient local caching. To this
    end we introduce a SIMD-optimized sparsity pattern that uses a hybrid
    storage format blending a \emph{packed row} (ELL) format for highly
    structures SIMD parallel regions with a more flexible \emph{compressed
    sparse row} (CSR) storage format for non-vectorized index regions.
  \item
    We report excellent weak and strong scaling of our implementation for
    both 2D and 3D problems, and demonstrate that our solver is able to
    tackle realistic 3D applications by computing a flow problem in 3D with
    about 1.8 billion gridpoints (totalling to about 8 billion spatial
    degrees of freedom).
  \item
    The main performance limitations of the solver are assessed,
    considering the mathematical model as fixed down to roundoff
    precision. Our analysis identifies which mathematical steps could be
    modified to further improve performance in the future. This analysis gives
    a guideline for performance optimization of a broader class of algorithms
    based on unstructured-grid stencil-based update formulas with complex data
    dependencies and heavy transcendental arithmetic.
    In addition, the analysis allows for predictions regarding the expected
    performance envelope on hardware with different characteristics than
    the present CPU-based architectures.
  \item
    A reference implementation of the solver is made available%
    \footnote{\url{https://doi.org/10.5281/zenodo.3924365}}
    that is based on the deal.II finite element library
    \cite{dealii91,dealiicanonical} and is freely available for the
    scientific community under an open source license.
\end{itemize}

The remainder of the paper is organized as follows. In
Section~\ref{sec:euler} we review the compressible Euler equations and
introduce important physical quantities.
In Section~\ref{sec:scheme} the solver is discussed in a concise, abstract
(mathematical) manner.
In particular, the invariant domain property of the solver is discussed in
Section~\ref{subsec:low-order} and the convex limiting paradigm is
introduced in Section~\ref{subsec:high-order}.
We summarize key design decisions of our
implementation in Section~\ref{sec:implementation} and report benchmark
results and explore algorithmic alternatives in
Section~\ref{sec:benchmarks}. We conclude in Section~\ref{sec:conclusion}
with a detailed discussion of possible further improvements that require
some mathematical reformulation.


\section{The Euler equations of gas dynamics}
\label{sec:euler}

Let $\Domain$ be an open polyhedral domain in $\R^d$, $d=1,2,3$. We
consider the compressible Euler equations in conservative form,
\begin{align}
  \label{eq:euler_equations}
  \partial_t\bu \,+\, \DIV \polf(\bu) = 0,
\end{align}
equipped with suitable initial conditions $\bu(\bx,0)=\bu_0$. Here, the
independent variables are $(\bx,t)\in \Domain\CROSS\R_+$ and the
vector $\bu:=(\rho,\bm,E)\tr \in \R^{d+2}$ describes the (dependent)
conserved quantities, the density $\rho$, the momentum $\bm$, and the total
energy $E$. The flux $\polf(\bu) \in \R^{(d+2)\times d}$ is given by
\begin{align}
 \polf(\bu)
 :=
 \begin{pmatrix}
   \bm \;,\; \bv\otimes \bm + p\polI_d \;,\; \bv(E+p)
 \end{pmatrix}\tr,
 \label{eq:flux}
\end{align}
where $\polI_d$ is the $d\CROSS d$ identity matrix, and $p$ is the pressure
that will be defined below. Starting from the vector $\bu$ of
conserved quantities we define a number of derived physical quantities.
The velocity of the fluid
particles is denoted $\bv:=\rho^{-1} \bm$ and $e := \rho^{-1}E - \frac12 \bv^2$ denotes the
specific internal energy. We call the quantity $\varepsilon := \rho
e= E -\frac12 \rho\bv^2$ internal energy. Here, we have used the
notation $\bv^2:= \|\bv\|^2$, where $\|\cdot\|$ is the Euclidean norm.

The pressure $p$ is defined by an equation of state
derived from a specific entropy $s(\rho,e)$
\cite{gp_first_order_2016, Harten_1983}. For the sake of simplicity we
limit the discussion in this paper to a polytropic ideal gas by setting
\begin{align*}
  s(\rho,e) - s_0 \;=\;
  \log\Big(e^{\frac{1}{\gamma-1}}\rho^{-1}\Big),
\end{align*}
where $\gamma$ is the ratio of specific heats that we set to
$\gamma=7\diagup5$. This implies that
\begin{align*}
  p \;:=\;
  -\rho^2\frac{\diff s}{\diff \rho}\Big(\frac{\diff s}{\diff e}\Big)^{-1}
  \;=\;
  (\gamma -1)\,\varepsilon.
\end{align*}
We also introduce the speed of sound
$c=\sqrt{\frac{\gamma\,p}{\rho}}$, as well as a scaled specific
entropy that will be used in the context of convex limiting,
\begin{equation}
  \label{eq:specific_entropy}
  \phi(\bu)
  \;:=\; \exp\big((\gamma-1)\,s(\rho, e(\bu))\big)
  \;=\; \varepsilon\,\rho^{-\gamma}.
\end{equation}
As a last preparatory step we introduce a Harten-type entropy
\cite[Eq.~2.10a]{Harten_1983},
\begin{equation}
  \eta(\bu)
  \;:=\; \big(\rho^2e\big)^{\frac{1}{\gamma+1}}
  \;=\; \big(\rho\varepsilon\big)^{\frac{1}{\gamma+1}}.
\end{equation}


\section{Second-order invariant-domain preserving Euler scheme}
\label{sec:scheme}

Before proceeding to the algorithmic details of our solver we summarize the
method in this section in a concise, mathematical manner. Our solver is
based on the convex-limiting technique pioneered by Guermond et
al.~\cite{gnpt_second_order_2018}. We refer the reader to
\cite{gp_first_order_2016, gp_fast_riemann_2016, gp_second_order_2018,
gnpt_second_order_2018} for a detailed derivation and
analysis of the respective building blocks.
We summarize and slightly adapt the algorithm here with the aim of
developing a scalable hybrid-parallelized solver that can utilize modern
hardware. In the following,
we introduce the underlying finite-element discretization, low- and
high-order update step, as well as necessary building blocks for
the final time stepping (Section~\ref{subsec:final-update}).

\subsection{Finite element discretization}
\label{subsec:finite-element}

Let $\TT_h$ be a partition of $\Omega$ into a shape-regular quadrilateral
or hexahedral mesh. We denote by $\big\{\varphi_i^h\big\}_i^\NN$ the
Lagrange basis of $\Qo(\TT_h)$, the space of piecewise linear,
bilinear, or trilinear finite elements on $\Omega$ ($d=1$, $2$, $3$). In
the following we will make use of two fundamental properties of the
Lagrange basis, the nonnegativity of the lumped mass matrix and a
partition of unity property, respectively,
\begin{align*}
  \int_\Omega\varphi_i^h\diff x > 0
  \quad
  \text{for }\;1\le i\le\NN,
  \qquad\qquad
  \sum_{i=1}^\NN\varphi_i^h(x) = 1
  \quad
  \text{for }x\in\Omega.
\end{align*}
Following the notation in \cite{gnpt_second_order_2018}, we
introduce a number of scalar and vector-valued matrix elements:
\begin{align}
  \begin{drcases}
    \begin{aligned}
      m_{ij} &\coloneqq \int_\Omega\varphi^h_i(x)\varphi^h_j(x)\diff x,
      &\qquad\qquad
      m_{i} &\coloneqq \int_\Omega\varphi^h_i(x)\diff x,
      \qquad
      \\[0.3em]
      \bc_{ij} &\coloneqq \int_\Omega\varphi^h_i(x)\GRAD\varphi^h_j(x)\diff x,
      &\qquad\qquad
      \bn_{ij} &\coloneqq \frac{\bc_{ij}}{\|\bc_{ij}\|},
      \qquad
      \\[0.3em]
      \beta_{ij} &\coloneqq \int_\Omega\GRAD\varphi^h_i(x)\SCAL\GRAD\varphi^h_j(x)
      \diff x,
      &\qquad\qquad
      b_{ij} &\coloneqq \delta_{ij} - \frac{m_{ij}}{m_j},
    \end{aligned}
  \end{drcases}
  \label{eq:matrices}
\end{align}
where $\delta_{ij}$ denotes Kronecker's delta. The matrices introduced in
\eqref{eq:matrices} only depend on the mesh and the particular choice of
the finite element basis. For a given index $i$, we introduce a stencil of
nonzero matrix entries
\begin{align*}
  \Ii\coloneqq\big\{1\le
  j\le\NN\;\big|\;\text{supp}\,(\varphi_i^h)\cap
  \text{supp}\,(\varphi_j^h)\not=\emptyset\big\}.
\end{align*}

\subsection{Efficient precomputation}

The solver algorithm discussed in the following consists of
nonlinear updates that are organized as loops over the stencil:
\\
\begin{algorithm}[H]
  \DontPrintSemicolon
  \For{$i=1$, \ldots, $\NN$}{
    \For{$j\in\Ii$}{
      \emph{(Nonlinear) computation involving quantities with indices $i$
      and $j$}
    }
  }
\end{algorithm}
\noindent
This is a \emph{stencil-centric} operation in contrast to the usual
\emph{cell-centric} loops typically encountered in finite element assembly
\cite{dealiicanonical}.
In order to achieve good performance, the first decision is whether the
matrices defined in \eqref{eq:matrices} should be recomputed ``on the fly''
in terms of a matrix-free approach, or whether it is more efficient to
precompute and store some matrices.
For low-order discretizations, matrix-free schemes based on fast integration
with sum factorization cannot amortize the work at quadrature points to a
sufficient number of degrees of freedom (dofs) on a cell, thus incurring a
substantial arithmetic overhead compared to matrix-based
schemes~\cite{Fischer_2020,Kronbichler_2012,Kronbichler_2019}. The overhead
is around 500 floating point operations per nonzero entry for tri-linear
polynomials in 3D using similar arguments as for the operator action
in \cite[Fig.~1]{Kronbichler_2012}. These computations are necessary because
some of the nonlinear
update steps specified below explicitly require the full value of the
$(i,j)$-th entry of the respective matrix. We point out that even
hierarchical, stencil-based matrix-free methods (such as
\cite{Bergen_2006}) will need to incorporate additional steps to treat
nonlinearities or deformed meshes
(we refer to \cite{Bauer_2018} for a possible approach). As
will be shown below, many steps are below the threshold of saturating
memory bandwidth in a matrix-based implementation for contemporary
hardware. Furthermore, a
reformulation of our algorithms
in terms of a cell-based loop, viz.\\
\begin{algorithm}[H]
  \DontPrintSemicolon
  \For{$T\in\TT_h$}{
    \For{$i$ with $\text{supp}\,\big(\varphi^h_i\big)\cap T\not=\emptyset$}{
      \For{$j$ with $\text{supp}\,\big(\varphi^h_j\big)\cap T\not=\emptyset$}{
        \ldots\phantom{A}
      }
    }
  }
\end{algorithm}
\noindent
  would necessitate additional communication from degrees of freedom
from different cells, which is better done before the time loop.
  Based on these considerations, the limiting resource identification
  underlying the roofline performance model~\cite{Williams_2009} suggests that
  the on-the-fly matrix-free computation would not relax the
  performance-limiting factor. Even though arithmetic intensity would be
  further increased, the application metric of the throughput in terms of
  points updated per second would decrease. Consequently, the most
  performance-beneficial setup is a stencil-based
  loop structure with pre-computed matrices.

  Starting from these considerations, we avoid all assembly operations during
  the time loop and precompute the three matrices $m_{ij}$, $\bc_{ij}$,
  $\beta_{ij}$. Note that each matrix contains unique information in terms of
  the shape functions. Furthermore, the frequent use of the diagonal matrix
  $m_i\delta_{ij}$ on the one hand and the low memory consumption on the other
  motivates to also store this matrix.  Given one layer of overlap in the mesh
  to the neighboring MPI ranks, the computation of those four matrices is
  completely local to
each MPI rank. Conversely, the matrices $\bn_{ij}$ and $b_{ij}$ are derived
on the fly from $\bc_{ij}$ and $m_{ij}$:
Matrix $\bn_{ij}$ is used in close proximity to $\bc_{ij}$,
  thus leading to a single division and three multiplications  of data present already in registers,
  which is cheaper than transferring three doubles through the memory
  hierarchy. The motivation for $b_{ij}$ is more subtle: The code below uses
  both $b_{ij}$ and $b_{ji}$ for the update; whereas $m_{ij}$ is symmetric,
  the matrix $b_{ij}$ is not. In the presence of caches, see the analysis
  below, it is hence cheaper to only load the symmetric entry $m_{ij}$ and the
  entries $1/m_{i}$ and $1/m_j$ derived from the diagonal mass matrix.  In addition, we
propose to precompute the inverse of the lumped mass matrix,
$(1/m_{i}\,\delta_{ij})$, in order to avoid divisions. We refer to the
detailed discussion in Section~\ref{sec:benchmarks}.

\subsection{Intermediate low-order update}
\label{subsec:low-order}

Given a snapshot $\big(\bU_i^n\big)_{1\le i\le\NN}$ of \emph{admissible}
states at time $t_n$ (this is to say that
$\rho(\bU_i^{n})>0$ and $\varepsilon(\bU_i^{n})>0$
) with an associated finite-element function $\bu_h^n=\sum_{i=1}^\NN
\bU_i^n\varphi^h_i$, our goal is to compute a new snapshot
$\big(\bU_i^{n+1}\big)_{1\le i\le\NN}$ consistent with the Euler
equations~\eqref{eq:euler_equations} such that the states maintain the
following crucial thermodynamical constraints
\begin{itemize}
  \item admissibility: positivity of density, $\rho(\bU_i^{n+1})>0$, and
    positivity of internal energy, $\varepsilon(\bU_i^{n+1})>0$,
  \item local minimum principle on specific entropy:
    $s(\bU_i^{n+1}) \ge \min_{j\in\Ii}\,s(U_j^{n})$.
\end{itemize}
The first algorithmic ingredient to achieve a high-order update obeying
above constraints is the computation of an intermediate low-order update
$\bU_i^{L,n+1}$ with a first-order \emph{graph viscosity} method
\cite{gp_first_order_2016}. The method is based on a guaranteed maximum
wavespeed estimate coming from an approximate Riemann solver
\cite{gp_fast_riemann_2016}. We construct an explicit update of the state
$u_h^n=\sum_{i=1}^\NN \bU_i^n\varphi^h_i$ at time $t_n$ for some new time
$t_{n+1}=t_n+\tau_n$ as follows:
\begin{align}
  \label{eq:low-order}
  \bU_i^{L,n+1} =
  \bU_i^{n} + \frac{\tau_n}{m_i}\sum_{j\in\Ii}\Big(-\polf(\bUnj)
  \SCAL\bc_{ij}
  +d_{ij}^{L,n}\big(\bUnj-\bUni\big)\Big).
\end{align}
Here, $d_{ij}^{L,n}$ is a graph viscosity given by
\begin{align}
  d_{ij}^{L,n} \;\coloneqq\;
  \max\,\Big(\tilde\lambda_{\max}(\bn_{ij},\bUni,\bUnj)\,|\bc_{ij}|
  \,,\,
  \tilde\lambda_{\max}(\bn_{ji},\bUnj,\bUni)\,|\bc_{ji}| \Big)
  \quad\text{for }i\not=j,
  \qquad
  d_{ii}^{L,n}=-\sum_{i\not=j\in\Ii}d_{ij}^{L,n},
\end{align}
where $\tilde\lambda_{\max}(\bn_{ij},\bUni,\bUnj)$ is a suitable upper
bound on the maximum wave speed in an associated one dimensional Riemann
problem \cite{gp_first_order_2016,gp_fast_riemann_2016}. The exact
definition of $\tilde\lambda_{\max}(\bn_{ij},\bUni,\bUnj)$ and description
of the approximate Riemann solver that is used in the computation is
postponed to Section~\ref{subsec:riemann-solver}. The time-step size is set
to
\begin{align}
  \tau_n = c_{\text{cfl}}\,\min_{1\le
  i\le\NN}\left(\frac{m_i}{-2\,d_{ii}^{L,n}}\right),
\end{align}
with a chosen constant $0<c_{\text{cfl}}\le1$. In preparation for the
high-order update with convex limiting, we rewrite the low-order update
\eqref{eq:low-order} as follows,
\begin{align}
  \label{eq:def-ubar}
  \bU_i^{L,n+1} =
  \bU_i^{L,n} + \frac{2\,\tau_n}{m_i}\sum_{j\in\Ii} d_{ij}^{L,n}\bUnij,
  \qquad
  \bUnij\;\coloneqq\;\frac12\big(\bUni+\bUnj\big)-\frac1{2\,d_{ij}^{L,n}}\big(
  \polf(\bUnj)-\polf(\bUni)\big)\SCAL\bc_{ij},
\end{align}
where we have used the identities $\sum_{j\in\Ii}\bc_{ij}=\bzero$, and
$\sum_{j\in\Ii}d^{L,n}_{ij}=0$.

\subsection{Intermediate high-order update}
\label{subsec:high-order}

We now introduce a formally high-order update that is \emph{entropy
consistent} and close to being \emph{invariant-domain preserving}
\cite{gnpt_second_order_2018}. The update is similar to the
low-order update \eqref{eq:low-order}, the only difference being that the
graph viscosity $d_{ij}^L$ of the low-order update is replaced by a
suitable $d_{ij}^H\le d_{ij}^L$ and the consistent mass matrix $m_{ij}$ is
used instead of the lumped mass matrix $m_i$,
\begin{align}
  \label{eq:high-order}
  \sum_{j\in\Ii} m_{ij}\big(
  \tilde\bU_j^{H,n+1} - \bU_j^{n}\big)
  \;=\;
  \tau_n\sum_{j\in\Ii}\Big(-\polf(\bUnj)
  \SCAL\bc_{ij}
  +d_{ij}^{H,n}\big(\bUnj-\bUni\big)\Big),
\end{align}
and where we set
\begin{align}
  d_{ij}^{H,n} \;\coloneqq\;
  d_{ij}^{L,n}\,\frac{\alpha_i^n+\alpha_j^n}{2}
  \quad\text{for }i\not=j,
  \qquad
  d_{ii}^{H,n}=-\sum_{i\not=j\in\Ii}d_{ij}^{H,n}.
\end{align}
Here, $\alpha_i^n$ denotes an \emph{indicator} given by a \emph{normalized
entropy viscosity ratio}. The precise definition and computation of
$\alpha_i^n$ is discussed in
Section~\ref{subsec:entropy-viscosity-commutator}. Solving for
$\tilde\bU_j^{H,n+1}$ given by \eqref{eq:high-order} involves inverting the
full mass matrix.
This is undesirable due to the high computational cost it incurs. Even with
a competitive preconditioner, solving \eqref{eq:high-order} can be as
expensive as the entire rest of the full (explicit) update step. We avoid
this issue and obtain a very efficient scheme by approximating the inverse of
the matrix with a Neumann series. This introduces a second-order consistency
error, which is however close to the underlying discretization error and much
smaller than the error caused by the lumped mass matrix. We start by
rewriting \eqref{eq:high-order} as follows
\begin{align}
  \label{eq:high-order-final}
  \sum_{j\in\Ii} \frac{m_{ij}}{m_{j}}
  \;
  \frac{m_{j}}{\tau_n}\big(
  \tilde\bU_j^{H,n+1} - \bU_j^{n}\big)
  \;=\;
  \bR^n_i,
  \qquad\text{with}\quad
  \bR^n_i\;\coloneqq\;
  \sum_{j\in\Ii}\Big(-\polf(\bUnj) \SCAL\bc_{ij}
  +d_{ij}^{H,n}\big(\bUnj-\bUni\big)\Big).
\end{align}
By expanding the inverse of the matrix $m_{ij}/m_j$ into a Neumann series
up to first order,
\begin{align*}
  \Big(\frac{m_{ij}}{m_j}\Big)^{-1} \,=\,
  \Big(\delta_{ij} -\big(\delta_{ij}-\frac{m_{ij}}{m_j}\big)\Big)^{-1}
  \;\approx\;\;
  \delta_{ij} +\big(\delta_{ij}-\frac{m_{ij}}{m_j}\big)
  \,=\,
  \delta_{ij} + b_{ij},
\end{align*}
we obtain
\begin{align*}
  \frac{m_{i}}{\tau_n}\big(
  \bU_i^{H,n+1} - \bU_i^{n}\big)
  \;=\;
  \bR^n_i + \sum_{j\in\Ii} \big(b_{ij} \bR^n_j - b_{ji} \bR^n_i\big).
\end{align*}
Here, we have used the fact that $\sum_{j\in\Ii}b_{ji}=0$ to add the second
term in the sum on the right hand side. By taking the difference of this
equation with equation \eqref{eq:low-order} that defines the low-order
update we obtain
\begin{align}
  \label{eq:def-p}
  \bU_i^{H,n+1} - \bU_i^{L,n+1}
  \;=\;
  \sum_{j\in\Ii} \lambda\,\bP^n_{ij},
  \quad
  \text{where}
  \quad
  \bP^n_{ij}\;\coloneqq\;
  \frac{\tau_n}{m_i\,\lambda}
  \left\{
  b_{ij} \bR^n_j - b_{ji} \bR^n_i
  + \big(d_{ij}^{H,n}-d_{ij}^{L,n}\big)\big(\bUnj-\bUni\big)
  \right\}.
\end{align}
In the above definition of $\bP^n_{ij}$ we have introduced an additional
scaling parameter, $\lambda\coloneqq 1 / (\text{card}\,\big(\Ii\big)-1)$, that
plays a crucial role in the convex limiting \cite{gnpt_second_order_2018}
discussed in Section~\ref{subsec:specific-entropy-limiter}.

\subsection{Full update step}
\label{subsec:final-update}

The actual update is now defined as follows. Given $\bUnij$, the low-order
update $\bU^{L,n+1}_i$, and $\bP^n_{ij}$ as defined in \eqref{eq:def-ubar}
and \eqref{eq:def-p}, the new state $\bU^{n+1}_i$ is constructed by means
of an iterative process \cite{gnpt_second_order_2018}: First, start by
setting
\begin{align*}
  \bU_i\;\leftarrow\;\bU^{L,n+1}_i,
  \qquad
  \bP_{ij}\;\leftarrow\;\bP^{n}_{ij}.
\end{align*}
Then, limiter bounds are computed and an update is performed:
\begin{align}
  \label{eq:iterative-limiter}
  \begin{drcases}
    l_{ij}\;=\;
    \min\Big(\text{limiter}\,\big(\bUnij\,;\, \bU_i,\, \bP_{ij}\big),\,
    \text{limiter}\,\big(\bUnij\,;\, \bU_j,\, \bP_{ij}\big) \Big),
    \qquad
    \\[0.3em]
    \bU_i\;\leftarrow\;\bU_i + \sum_{j\in\Ii} \lambda\,l_{ij}\,\bP_{ij},
    \qquad
    \bP_{ij}\;\leftarrow\;(1-l_{ij})\bP_{ij}.
  \end{drcases}
\end{align}
The discussion of the limiter function is deferred to
Section~\ref{subsec:specific-entropy-limiter}. For reasons of stability, at
least two passes of update step \eqref{eq:iterative-limiter} are performed
before accepting the current value by setting
$\bU^{n+1}_i\,\coloneqq\,\bU_i$. For the convenience of the reader the full
update procedure is summarized as pseudo code in Alg.~\ref{alg:euler}.
\begin{algorithm}[tbp]
  \DontPrintSemicolon
  \SetKwProg{Euler}{euler\_step}{}{end}
  \Euler{}{
    \tcp*[l]{Step 0: precompute entropies (see Section~\ref{sec:benchmarks})}
    \tcp*[l]{Step 1: compute off-diagonal $d_{ij}^{L,n}$ and $\alpha_i$:}
    \For{$i=1$, \ldots, $\NN$}{
      \texttt{indicator.reset($\bU_i^n$)}\;
      \For{$j\in\Ii$, $j>i$}{
        $d_{ij}^{L,n}\;\leftarrow\;
        \max\,\Big(\tilde\lambda_{\max}(\bn_{ij},\bUni,\bUnj)\,|\bc_{ij}|\,,\,
        \tilde\lambda_{\max}(\bn_{ji},\bUnj,\bUni)\,|\bc_{ji}| \Big)$\;
        \texttt{indicator.accumulate($\bU_j^n$, $\bc_{ij}$, $\beta_{ij}$)}
      }
      $\alpha_i\;\leftarrow\;\text{\texttt{indicator.result()}}$
    }

    \vspace{0.3em}\tcp*[l]{Step 2: fill lower-diagonal part and compute $d_{ii}^{L,n}$ and $\tau_n$:}
    $\tau_n\;\leftarrow\;+\infty$\;
    \For{$i=1$, \ldots, $\NN$}{
      \For{$j\in\Ii$, $j<i$}{
        $d_{ij}^{L,n}\;\leftarrow\;d_{ji}^{L,n}$\;
      }
      $d_{ii}^{L,n}\;\leftarrow\;-\sum_{j\in\Ii,j\not=i}d_{ij}^{L,n}$
      \quad;\quad
      $\tau_n\;\leftarrow\;\min\Big(\tau_n\,,\,-c_{\text{cfl}}\frac{m_i}{2d_{ii}^{L,n}}\Big)$\;
    }

    \vspace{0.3em}\tcp*[l]{Step 3: low-order update, compute $\bR_i$ and accumulate limiter bounds}
    \For{$i=1$, \ldots, $\NN$}{
      \For{$j\in\Ii$}{
        $d_{ij}^{H,n} \;\leftarrow\; d_{ij}^{L,n}\,\frac{\alpha_i^n+\alpha_j^n}{2}$
        \qquad;\qquad
        $\bR^n_i \;\leftarrow\; \bR^n_i \;-\; \polf_j \SCAL\bc_{ij} + d_{ij}^{H,n}\big(\bUnj-\bUni\big)$\;
        $\bUnij \;\leftarrow\; \frac12\big(\bUni+\bUnj\big)-\frac1{2\,d_{ij}^{L,n}}\big(\polf_j-\polf_i\big)\SCAL\bc_{ij}$
        \;
        $\bU^{n+1}_i \;\leftarrow\; \frac{2\,\tau_n}{m_i}\,d^{L,n}_{ij}\bUnij$\;
        \texttt{limiter.accumulate\_bounds($\bU_i$, $\bU_j$, $\bUnij$)}\;
      }
      $\texttt{bounds}_i\;\leftarrow\;\text{\texttt{limiter.bounds()}}$
    }

    \vspace{0.3em}\tcp*[l]{Step 4: compute $\bP_{ij}$ and $l_{ij}$:}
    \For{$i=1$, \ldots, $\NN$}{
      \For{$j\in\Ii$}{
        $\bP_{ij}\;\leftarrow\;\frac{\tau_n}{\lambda m_i}\Big(\big(d_{ij}^{H,n}-d_{ij}^{L,n}\big)\big(\bUnj-\bUni\big)+b_{ij}\bR_j-b_{ji}\bR_i\Big)$\;
        $l_{ij}\;\leftarrow\;\texttt{limiter.compute($\bU_i^{n+1}$\,,\,$\bP_{ij}$\,,\,bounds$_i$)}$\;
      }
    }

    \vspace{0.3em}
    \For{pass $=\,1$, \ldots, number of limiter passes}{
      \tcp*[l]{Step 5, 6, \ldots: high-order update and recompute $l_{ij}$:}
      \For{$i=1$, \ldots, $\NN$}{
        \For{$j\in\Ii$}{
          $\bU^{n+1}_i \;\leftarrow\; \bU^{n+1}_i + \lambda
          \min(l_{ij},l_{ji}) \bP_{ij}^n$\;
        }
        \If{not last round}{
          \For{$j\in\Ii$}{
            $\bP_{ij}\;\leftarrow\; \big(1-\min(l_{ij},l_{ji})\big)\,\bP_{ij}$\;
            $l_{ij}\;\leftarrow\;\texttt{limiter.compute($\bU_i^{n+1}$\,,\,$\bP_{ij}$\,,\,bounds$_i$)}$\;
          }
        }
      }
    }
  }
  \caption{High-order forward Euler step. The indicator and limiter are
    discussed in Section~\ref{subsec:entropy-viscosity-commutator} and
    \ref{subsec:specific-entropy-limiter}. The $\tilde\lambda_{\text{max}}$
    values are computed with an approximate Riemann solver discussed in
    Section~\ref{subsec:riemann-solver}.}
  \label{alg:euler}
\end{algorithm}

\subsection{Strong stability preserving Runge-Kutta scheme}
\label{subsec:ssp-rk}

The update process described so far is second order in space but only first
order in time. In order to obtain a scheme that is also high-order in time,
we combine the update process with a third-order strong stability
preserving (SSP) Runge-Kutta scheme \cite{Shu_1988}. More precisely, let
$\tau_n$, $\bU_i^{n+1,(1)}$ denote the computed time-step size and the
computed update of iterative process $\eqref{eq:iterative-limiter}$. We
then repeat the update step described above in order to compute a second
intermediate state $\bU_i^{n+1,(2)}$ and the actual update $\bU_i^{n+1}$ by
replacing the original state $\bU_i^n$ by $\bU_i^{n+1,(1)}$, and
$\bU_i^{n+1,(2)}$, (while keeping the time-step size $\tau_n$ fixed) and by
scaling the result,
\begin{align}
  \label{eq:ssprk3}
  \begin{drcases}
    \begin{aligned}
      \tau_n,\;\bU_i^{n+1,(1)}\quad &\leftarrow\quad
      \text{\texttt{euler\_step}}\,\big(\,\bUni\,\big),
      \\[0.3em]
      \bU_i^{n+1,(2)}\quad &\leftarrow\quad
      \frac{3}{4}\,\bUni \;+\; \frac{1}{4}\,
      \text{\texttt{euler\_step}}\,\Big(\tau_n,\;\bU_i^{n+1,(1)}\Big),
      \\[0.3em]
      \bU_i^{n+1}\quad &\leftarrow\quad
      \frac{1}{3}\,\bUni \;+\; \frac{2}{3}\,
      \text{\texttt{euler\_step}}\,\Big(\tau_n,\;\bU_i^{n+1,(2)}\Big).
      \quad
    \end{aligned}
  \end{drcases}
\end{align}

\subsection{Approximate Riemann solver}
\label{subsec:riemann-solver}

For constructing the \emph{graph viscosity}
\begin{align*}
  d_{ij}^{L,n} \;=\;
  \max\,\Big(\tilde\lambda_{\max}(\bn_{ij},\bUni,\bUnj)\,|\bc_{ij}|
  \,,\,
  \tilde\lambda_{\max}(\bn_{ji},\bUnj,\bUni)\,|\bc_{ji}| \Big),
\end{align*}
sharp upper bounds on the maximal wave speed
$\tilde\lambda_{\max}(\bn_{ij},\bUni,\bUnj)$ of the associated 1D Riemann
problem can be computed with fast, approximate Riemann solvers
\cite{gp_fast_riemann_2016}. For our purpose, however, the low-order
articifical viscosity $d_{ij}^{L,n}$ is allowed to be overestimated to a
certain extent without degrading the performance of the second-order
scheme. We thus only use an inexpensive guaranteed upper bound on the
maximum wave speed by means of a \emph{two-rarefaction} approximation
\cite{gp_fast_riemann_2016} (and that would ordinarily used as a starting
point for a quadratic Newton iteration \cite{gp_fast_riemann_2016}). This
choice has the added benefit that the approximate Riemann solver can also
be efficiently SIMD parallelized as will be discussed in
Sections~\ref{sec:implementation} and \ref{sec:benchmarks}. For a given
state $\bU$ and direction $\bn_{ij}$, a projected 1D state is defined as
follows
\begin{align*}
  \tilde\rho\;\coloneqq\;\rho,
  \qquad
  \tilde m\;\coloneqq\;\bn_{ij}\SCAL\bm,
  \qquad
  \tilde E \;\coloneqq\; E - \frac{1}{2\,\rho}\big\|\bm-\tilde
  m\,\bn_{ij}\|_{l^2}^2.
\end{align*}
We now introduce two quantities of characteristic propagation speeds that
depend on a pressure $p^\ast$ and either the $\bUni$ or $\bUnj$ state
\cite{gp_fast_riemann_2016},
\begin{align*}
  \lambda_-^1(\bUni,p^\ast)
  \,\coloneqq\,\tilde u^n_i \;-\; \tilde
  c^n_i\,\sqrt{1+\frac{\gamma+1}{2\,\gamma}\left[\frac{p^\ast-\tilde
  p^n_i}{\tilde p^n_i}\right]_{\text{pos}}},
  \qquad
  \lambda_+^3(\bUnj,p^\ast)
  \,\coloneqq\,\tilde u^n_j \;+\; \tilde
  c^n_j\,\sqrt{1+\frac{\gamma+1}{2\,\gamma}\left[\frac{p^\ast-\tilde
  p^n_j}{\tilde p^n_j}\right]_{\text{pos}}},
\end{align*}
where we have used the symbol
$\left[\,x\,\right]_{\text{pos}}=\frac{|x|+x}{2}$, and where the derived
quantities $\tilde c$ and $\tilde p$ are computed from the corresponding
projected 1D states. A two-rarefaction pressure $\tilde
p^\ast(\bUni,\bUnj)$ is given by
\begin{align*}
  \tilde p^\ast(\bUni,\bUnj)
  = \tilde p_j\,
  \left(
    \frac{\tilde c_i + \tilde c_j - \frac{\gamma-1}{2}\big(\tilde
    u_j-\tilde u_i\big)}{\tilde c_i\,\left(\frac{\tilde p_i}{\tilde
    p_j}\right)^{-\frac{\gamma-1}{2\,\gamma}}+\tilde c_j}
  \right)^{\frac{2\,\gamma}{\gamma-1}},
\end{align*}
and a monotone increasing and concave down function
\cite{gp_fast_riemann_2016} is constructed as follows
\begin{align*}
  \psi(p)\;\coloneqq\;f(\bUni,p)+f(\bUnj,p)+\tilde u_j - \tilde u_i,
  \qquad
  f(\bU,p)\;\coloneqq\;
  \begin{cases}
    \begin{aligned}
      \frac{\sqrt{2}\,(p-\tilde p)}
      {\sqrt{\tilde\rho\big[(\gamma+1)\,p+(\gamma-1)\,\tilde p\big]}},
      &\quad &\text{if }{p\ge\tilde p},
      \\[0.3em]
      \left[\left(p/\tilde p\right)^{\frac{\gamma-1}{2\,\gamma}} -
      1\right]\,\frac{2\,\tilde c}{\gamma-1},
      &\quad &\text{otherwise}.
    \end{aligned}
  \end{cases}
\end{align*}
By using these ingredients, the wave speed estimate is constructed
as follows,
\begin{align*}
  \tilde\lambda_{\text{max}}=\max\left(
  \big[\lambda^1_-(\bUni,p^\ast)\big]_{\text{neg}}\,,\,
  \big[\lambda^3_+(\bUnj,p^\ast)\big]_{\text{pos}}
  \right),
\end{align*}
and where
\begin{align*}
  p^\ast \;\coloneqq\;
  \begin{cases}
    \begin{aligned}
      &\tilde p^\ast(\bUni,\bUnj)
      &\quad
      &\text{if } \psi(p_{\max}) < 0,
      \\
      &\min(p_{\max}, \tilde p^\ast(\bUni,\bUnj))
      &\quad
      &\text{otherwise}.
    \end{aligned}
  \end{cases}
\end{align*}
with the definitions $p_{\min}=\min(\tilde p_i,\tilde p_j)$ and
$p_{\max}=\max(\tilde p_i,\tilde p_j)$.

\subsection{Entropy viscosity commutator}
\label{subsec:entropy-viscosity-commutator}
The indicator used for constructing the high-order solver is an
entropy-viscosity commutator as described in \cite{gpp_2011, gnpt_second_order_2018}.
We choose the Harten entropy $\eta$ as described in
Section~\ref{sec:euler}. Let $\eta'$ denote its derivative with respect to
the state variables:
\begin{align*}
  \eta'(\bU)\;=\;\frac{(\rho\,\varepsilon)^{-\gamma/(\gamma + 1)}}{\gamma+1}
  \begin{pmatrix}
    E \\ -\bm \\ \rho
  \end{pmatrix}.
\end{align*}
With the help of the two quantities
\begin{align*}
  a_i^n \;\coloneqq\;
  \sum_{j\in\Ii}\left(\frac{\eta(\bU_j^n)}{\rho_j^n}
  -\frac{\eta(\bU_i^n)}{\rho_i^n}\right)\,\bm_j^n\SCAL\bc_{ij},
  \qquad
  \bb_i^n \;\coloneqq\;
  \sum_{j\in\Ii}\left(\polf(\bU_j^n)-\polf(\bU_i^n)\right)\SCAL\bc_{ij},
\end{align*}
the normalized entropy viscosity ratio $\alpha_i^n$ for the state $\bU_i^n$
is now constructed as follows:
\begin{align*}
  \alpha_i^n\;=\;\frac{N_i^n}{D_i^n},
  \quad
  N_i^n\;\coloneqq\;\left|a_i^n- \eta'(\bUni)\SCAL\bb_i^n
  +\frac{\eta(\bUni)}{\rho_i^n}\big(\bb_i^n\big)_1\right|,
  \quad
  D_i^n\;\coloneqq\;\left|a_i^n\right| +
  \sum_{k=1}^{d+1}\left|\big(\eta'(\bUni)\big)_k-
  \delta_{1k}\frac{\eta(\bUni)}{\rho_i^n}\right|
  \,\left|\big(\bb_i^n\big)_k\right|,
\end{align*}
where $\big(\,.\,\big)_k$ denotes the $k$-th component of a vector and
$\delta_{ij}$ is Kronecker's delta.

\subsection{Convex limiting on specific entropy}
\label{subsec:specific-entropy-limiter}

The starting point of our discussion of the limiting process is Equation
\eqref{eq:def-p}, viz.,
\begin{align*}
  \bU_i^{H,n+1} \;=\; \bU_i^{L,n+1} \,+\,
  \sum_{j\in\Ii}\lambda\bP_{ij}^n.
\end{align*}
We recall that $\bU_i^{L,n+1}$ is the intermediate low-order update that
ensures that all thermodynamical constraints are maintained (see
Section~\ref{subsec:low-order}). Unfortunately, the high-order
update $\bU_i^{H,n+1}$ is invariant domain violating and cannot be used
immediately. We thus limit the high-order update by introducing
$l_{ij}\in[0,1]$,
\begin{align}
  \tilde{\bU}_i \;=\; \bU_i^{L,n+1} \,+\,
  \sum_{j\in\Ii}\lambda l_{ij}\bP_{ij}^n.
  \label{eq:temp}
\end{align}
such that $l_{ij}=l_{ji}$ (to ensure conservation) and such that
$\tilde{\bU}_i$ maintain all stated thermodynamical constraints. Equation
\eqref{eq:temp} allows to break down the search for the factors $(l_{ij})$
into successive one-dimensional root finding problems that can be solved
very efficiently:
\begin{align*}
  \text{max}\,!\;\;{\tilde l_{ij}\in[0,1]}\quad\text{s.\,t.}\quad
  \bU_i+l_{ij}\bP_{ij}\;\;\text{maintains thermodynamical constraints.}
\end{align*}
A key observation is the fact that the $\tilde l_{ij}$ found in that way
have the property that the combined update $\eqref{eq:temp}$ obeys the
thermodynamical constraints as well \cite{gnpt_second_order_2018}. The
downside of this approach, however, is the fact that the factors are not
necessarily optimal. This can be improved by repeating the limiting step a
second time (as outlined in Section~\ref{subsec:final-update}).

For a given index $i$ we first define local bounds for the density and
specific entropy (the computation of these correspond to the
\texttt{limiter.accumulate\_bounds} call in Alg.~\ref{alg:euler}):
\begin{align*}
  \begin{cases}
    \begin{aligned}
      \rho_{\text{min}} \;&\coloneqq\; \min_{j\in\Ii}\rho\,(\bUnij),
      \\
      \rho_{\text{max}} \;&\coloneqq\; \max_{j\in\Ii}\rho\,(\bUnij),
      \\
      \phi_{\text{min}} \;&\coloneqq\; \min_{j\in\Ii}\,\phi\,(\bU_j).
    \end{aligned}
  \end{cases}
\end{align*}
\begin{remark}
  These bounds can be relaxed in order to obtain optimal 2nd-order
  convergence rates for smooth manufactured solutions, we refer the reader
  to \cite[Sec.~4.7]{gnpt_second_order_2018}. The relaxation procedure is
  implemented in our accompanying source code. For the sake of simplicity,
  however, we refrain from discussing the relaxation procedure.
\end{remark}

Given above bounds and an update \emph{direction} $\bP_{ij}$ one can now
determine a candidate $\tilde l_{ij}$ by computing
\begin{align*}
  \tilde l_{ij} = \max_{l\,\in\,[0,1]}
  \,\Big\{\rho_{\text{min}}\,\le\,\rho\,(\bU_i +\tilde l_{ij}\bP_{ij})
  \,\le\,\rho_{\text{max}},\quad
  \phi_{\text{min}}\,\le\,\phi\,(\bU_{i}+\tilde l_{ij}\bP_{ij})\Big\}.
\end{align*}
Algorithmically this is accomplished as follows: We first determine an
interval $[t_L,t_R]$ by setting $t_L=0$ and choosing $t_R\le 1$ ensuring
the bounds on the density \cite{gnpt_second_order_2018}. We then perform a
\emph{quadratic} Newton iteration \cite{gp_fast_riemann_2016} solving for
the root of a 3-convex function \cite{gp_fast_riemann_2016}
\begin{align*}
  \Psi(\bU)\;=\;\rho^{\gamma+1}(\bU)\,\big(\phi(\bU)-\phi_{\text{min}}\big).
\end{align*}
We note that by definition of $\Psi$ the condition $\Psi(\bU)\ge 0$ ensures
that the local minimum principle on the specific entropy is fulfilled. In
addition, $\Psi(\bU)\ge 0 $ also guarantees positivity of the internal
energy by virtue of equation~\eqref{eq:specific_entropy}. Initially we have
$\Psi(\bU_i+t_L\bP_{ij})\ge 0$, i.\,e. the factor $t_L$ is an admissible
limiter value. On the other hand, $t_R$ might be inadmissible, i.\,e.
$\Psi(\bU_i+t_L\bP_{ij}) < 0$. The quadratic Newton step updates the bounds
$t_L$ and $t_R$ simulatenously maintaining the property
$\Psi(\bU_i+t_L\bP_{ij})\ge 0 \ge \Psi(\bU_i+t_L\bP_{ij})$. The limiter
step is oulined in detail in Alg.~\ref{alg:limiter}.
\begin{algorithm}[tbp]
  \DontPrintSemicolon
  \SetKwProg{Limiter}{limiter.compute}{}{end}
  \Limiter{\texttt{($\bU_i$, $\bP_{ij}$, bounds)}}{
    \tcp*[l]{Ensure positivity of the density $\rho$:}
    $t_L \;\leftarrow\; 0$\;\vspace{0.3em}
    $t_R \;\leftarrow\;
    \begin{cases}
      \begin{aligned}
        1 \qquad &\text{if }
        \rho\,(\bU_i+t_R\bP_{ij})\,\le\,\rho_{\text{max}},
        \\
        \frac{\big|\rho_{\text{max}}-\rho\,(\bU_i)\big|}{\big|\rho\,(\bP_{ij})\big|}
        \qquad &\text{else.}
      \end{aligned}
    \end{cases}$\;\vspace{0.3em}
    $t_R \;\leftarrow\;
    \begin{cases}
      \begin{aligned}
        t_R \qquad &\text{if }
        \rho\,(\bU_i+t_R\bP_{ij})\,\ge\,\rho_{\text{min}},
        \\
        \frac{\big|\rho_{\text{min}}-\rho\,(\bU_i)\big|}{\big|\rho\,(\bP_{ij})\big|}
        \qquad &\text{else.}
      \end{aligned}
    \end{cases}$\;
    \vspace{0.3em}
    \tcp*[l]{Perform quadratic Newton update:}
    \For{step $=\,1$, \ldots, max number of Newton steps}{
      $\Psi_R \;\leftarrow\; \Psi(\bU_i+t_RP_{ij})$\;
      \tcp*[l]{If $\Psi_R\ge0$, then $t_R$ is already a good state, close interval:}
      $t_L \;\leftarrow\;
      \begin{cases}
        \begin{aligned}
          t_R \qquad &\text{if }
          \Psi_R\ge0,
          \\
          t_L
          \qquad &\text{else.}
        \end{aligned}
      \end{cases}$\;
      \If{$\Psi_R\ge 0$}{
        \tcp*[l]{$t_R$ is already a good state, exit for loop}
        {\bfseries break}\;
      }
      $\Psi_L \;\leftarrow\; \Psi(\bU_i+t_LP_{ij})$\;
      \If{$\Psi_L\le \text{TOL}$}{
        \tcp*[l]{within a preset tolerance $t_L$ is a root of $\Psi$, exit for loop}
        {\bfseries break}\;
      }
      $\text{d}\Psi_L \;\leftarrow\;
      \frac{\text{d}\Psi}{\text{d}t}(\bU_i+t\bP_{ij})\,\big|_{t=t_R}$\;
      $\text{d}\Psi_R \;\leftarrow\;
      \frac{\text{d}\Psi}{\text{d}t}(\bU_i+t\bP_{ij})\,\big|_{t=t_L}$\;
      $[t_L,t_R] \;\leftarrow\;$\texttt{quadratic\_newton\_step($t_L$, $t_R$, $\Psi_L$, $\Psi_R$, $\text{d}\Psi_L$, $\text{d}\Psi_R$)}.
    }
    \vspace{0.3em}
    \tcp*[l]{Accept $t_L$ as limiter bound:}
    $\tilde l_{ij} \;\leftarrow\; t_L$
  }
  \caption{%
    The convex limiting procedure. The unusual control flow in the
    algorithm ensures a straight-forward SIMD vectorization; see
    Section~\ref{subse:vectorization}.}
  \label{alg:limiter}
\end{algorithm}

\begin{algorithm}[tbp]
  \DontPrintSemicolon
  \SetKwProg{Newton}{quadratic\_newton\_step}{}{end}
  \Newton{\texttt{($t_L$, $t_R$, $\Psi_L$, $\Psi_R$, $\text{d}\Psi_L$, $\text{d}\Psi_R$, sign)}}{
    scaling $\;\leftarrow\; 1\,/\,(t_R - t_L + \text{eps})$

    $d_{11}  \;\leftarrow\; \text{d}\Psi_L$
    \quad;\quad
    $d_{12}  \;\leftarrow\; (\Psi_R - \Psi_L) \cdot \texttt{scaling}$
    \quad;\quad
    $d_{22}  \;\leftarrow\; \text{d}\Psi_R$\;
    $d_{112} \;\leftarrow\; (d_{12} - d_{11}) \cdot \texttt{scaling}$
    \quad;\quad
    $d_{122} \;\leftarrow\; (d_{22} - d_{12}) \cdot \texttt{scaling}$\;
    \vspace{0.3em}

    $\Lambda_L \;\leftarrow\; \big(\text{d}\Psi_L\big)^2 - 4\,\Psi_L\,d_{112}$
    \quad;\quad
    $\Lambda_R \;\leftarrow\; \big(\text{d}\Psi_R\big)^2 - 4\,\Psi_R\,d_{122}$\;
    \vspace{0.3em}

    $t_L \;\leftarrow\; t_L \;-\;\frac{2\,\Psi_L}{\text{d}\Psi_L\;+\;\texttt{sign}\,\sqrt{\Lambda_L}}$
    \quad;\quad
    $t_R \;\leftarrow\; t_R \;-\;\frac{2\,\Psi_R}{\text{d}\Psi_R\;+\;\texttt{sign}\,\sqrt{\Lambda_R}}$\;
    \vspace{0.3em}
    {\bfseries return} $[t_L,t_R]$
  }
  \vspace{0.3em}
  \caption{%
    Quadratic Newton step with divided differences. The input function
    $\Psi$ has to be 3-convex, i.\,e. the third derivative of $\Psi$ must
    be nonzero with a fixed positive or negative \texttt{sign}. (An actual
    implementation of the quadratic Newton scheme should take numerical
    round-off errors into account which requires additional safeguards not
    discussed here.)}
  \label{alg:newton}
\end{algorithm}


\section{Implementation}
\label{sec:implementation}

In this section we discuss the central implementation details of
the algorithm introduced in Section \ref{sec:scheme}. Particular emphasis
is on the local index handling and SIMD-optimized data
structures.

\subsection{Distributed and shared memory parallelism}
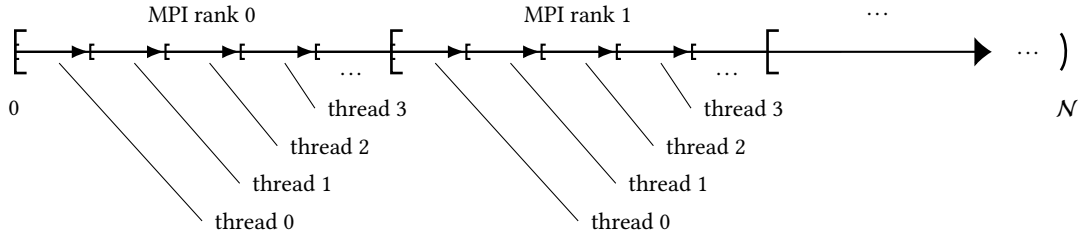
\begin{figure}
  \begin{tikzpicture}
    \node at ( 0.0, -0.75) {$0$};
    \node at ( 14.0, -0.75) {$\NN$};
    \draw[>={Latex[width=4mm]},->, thick] (0, 0) -- (13, 0);
    \draw[>={Bracket[width=6mm,line width=1pt,length=1.5mm]},<-] (0, 0) -- (1, 0);
    \node at ( 2.5, 0.5) {MPI rank 0};
    \draw[{Bracket}-{Latex}, thick] (0.0, 0.0) -- (1.0, 0.0);
    \draw[{Bracket}-{Latex}, thick] (1.0, 0.0) -- (2.0, 0.0);
    \draw[{Bracket}-{Latex}, thick] (2.0, 0.0) -- (3.0, 0.0);
    \draw[{Bracket}-{Latex}, thick] (3.0, 0.0) -- (4.0, 0.0);
    \draw[{Bracket}-, thick] (4.0, 0.0) -- (5.0, 0.0);
    \node at (4.5, -0.30) {\ldots};
    \draw[-] (4.0, -0.75) -- (3.6, -0.15);
    \node at  (4.7, -0.75) {thread 3};
    \draw[-] (3.5, -1.25) -- (2.6, -0.15);
    \node at  (4.2, -1.25) {thread 2};
    \draw[-] (3.0, -1.75) -- (1.6, -0.15);
    \node at  (3.7, -1.75) {thread 1};
    \draw[-] (2.5, -2.25) -- (0.6, -0.15);
    \node at  (3.2, -2.25) {thread 0};
    \draw[>={Bracket[width=6mm,line width=1pt,length=1.5mm]},<-] (5, 0) -- (6, 0);
    \node at ( 7.5, 0.5) {MPI rank 1};
    \draw[{Bracket}-{Latex}, thick] (5.0, 0.0) -- (6.0, 0.0);
    \draw[{Bracket}-{Latex}, thick] (6.0, 0.0) -- (7.0, 0.0);
    \draw[{Bracket}-{Latex}, thick] (7.0, 0.0) -- (8.0, 0.0);
    \draw[{Bracket}-{Latex}, thick] (8.0, 0.0) -- (9.0, 0.0);
    \draw[{Bracket}-, thick] (9.0, 0.0) -- (10.0, 0.0);
    \node at (9.5, -0.30) {\ldots};
    \draw[-] (9.0, -0.75) -- (8.6, -0.15);
    \node at  (9.7, -0.75) {thread 3};
    \draw[-] (8.5, -1.25) -- (7.6, -0.15);
    \node at  (9.2, -1.25) {thread 2};
    \draw[-] (8.0, -1.75) -- (6.6, -0.15);
    \node at  (8.7, -1.75) {thread 1};
    \draw[-] (7.5, -2.25) -- (5.6, -0.15);
    \node at  (8.2, -2.25) {thread 0};
    \draw[>={Bracket[width=6mm,line width=1pt,length=1.5mm]},<-] (10, 0) -- (11, 0);
    \node at (11.5, 0.5) {\ldots};
    \node at (13.5, -0.02) {\ldots};
    \draw[>={Parenthesis[width=6mm,line width=1pt,length=1.5mm]},->] (13.99, 0) -- (14, 0);
  \end{tikzpicture}
  \caption{%
    Hybrid process and thread parallelism: The index range $\NN$ is divided
    into contiguous ranges distributed over all MPI ranks, that in turn spawn
    threads subdividing the index range further.}
  \label{fig:process-thread-parallel}
\end{figure}
\begin{figure}
  \begin{tikzpicture}
    \draw[>={Latex[width=4mm]},->, thick] (-1, -0.5) -- (12, -0.5);
    \node at ( 0.0, 0) {Step 1};
    \node at ( 2.0, 0) {Step 2};
    \node at ( 4.0, 0) {Step 3};
    \node at ( 6.0, 0) {Step 4};
    \node at ( 8.0, 0) {Step 5};
    \node at (10.0, 0) {Step 6};
    \draw [fill] (0,-0.5) circle [radius=2pt];
    \draw [fill] (2,-0.5) circle [radius=2pt];
    \draw [fill] (4,-0.5) circle [radius=2pt];
    \draw [fill] (6,-0.5) circle [radius=2pt];
    \draw [fill] (8,-0.5) circle [radius=2pt];
    \draw [fill] (10,-0.5) circle [radius=2pt];
    \draw (2.75,0.0) -- (2.75,-1.5);
    \draw (4.75,0.0) -- (4.75,-1.5);
    \draw (6.75,0.0) -- (6.75,-1.5);
    \draw (8.75,0.0) -- (8.75,-1.5);
    \draw (10.75,0.0) -- (10.75,-1.5);
    \node[right, align=left] at (2.75, -1.38) {sync $\alpha_i$\\barrier $\tau_{\text{max}}$};
    \node[right, align=left] at (4.75, -1.2) {sync $\bR_i$};
    \node[right, align=left] at (6.75, -1.2) {sync $l_{ij}$};
    \node[right, align=left] at (8.75, -1.2) {sync $l_{ij}$};
    \node[right, align=left] at (10.75, -1.2){sync $\bU_i$};
  \end{tikzpicture}
  \caption{%
    MPI synchronization and barriers for Alg.~\ref{alg:euler} for the
    typical case of two limiter passes. During the execution of the forward
    Euler step (Alg.~\ref{alg:euler}) the $\alpha_i$, $\bR_i$ and $\bU_i$
    vectors and the $l_{ij}$ matrix have to be synchronized over MPI ranks:
    This incurs some MPI communication and forces an individual MPI rank to
    wait until all necessary data is received. The computation of the
    maximal admissible step size, $\tau_{\text{max}}$, requires an MPI
    Allreduce operation and thus incurs an MPI barrier after step 2 during
    which all MPI ranks have to wait for each other such that
    $\tau_{\text{max}}$ can be computed.}
  \label{fig:mpi-synchronization}
\end{figure}
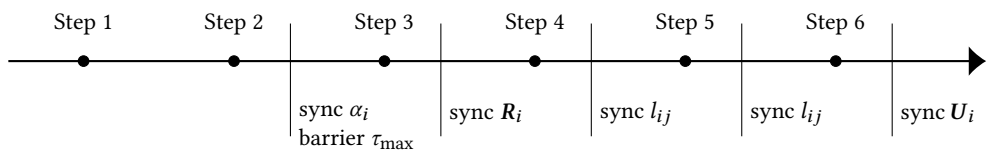

All building blocks of Alg.~\ref{alg:euler} are loops over
the stencil:\\
\begin{algorithm}[H]
  \DontPrintSemicolon
  \For{$i=1$, \ldots, $\NN$}{
    \For{$j\in\Ii$}{
      \emph{Computation involving index $i$ and $j$.}
    }
  }
\end{algorithm}
\noindent Since the computed updates to different indices $i$ are independent,
the parallelization with MPI and threads is straight-forward: First,
partition the set $\NN$ of indices among the participating MPI ranks. Then,
the local index ranges can be traversed in parallel by a number of workers;
see Fig.~\ref{fig:process-thread-parallel}.
Introducing shared-memory thread parallelism into the algorithm requires
only minimal modifications, mainly introducing thread-local temporary
memory and parallel for loops.
We have based our implementation on OpenMP~\cite{openmp45} because it is
readily supported by current C++ compilers.

In contrast, for distributed-memory parallelism we have to communicate
information contained in vector entries associated to the columns $\Ii$
between participating MPI ranks. We will comment on the precise handling of
such export and import indices in Section~\ref{subse:local_index}. For the
time being we observe that Alg.~\ref{alg:euler} is organized such that an
individual step computes a quantity (for example $\bR_i$ in step 3) that in
turn is needed in a subsequent step when looping over the stencil (for
example, $\bR_j$ for $j\in\Ii$ is used in step 4). Hence, all the values
$\bR_j$ need to be ready before proceeding with the next step, including
those values computed by another MPI rank which must be exchanged by a
suitable export step. Due to the arithmetic intensity in these steps as
explored in Section~\ref{sec:benchmarks} below, we consider global loops
for each of the steps. Wavefront diamond blocking away from the MPI
processor boundary would be possible to increase data locality between the
steps for the case of lower arithmetic
loads~\cite{Malas_2015,Malas_2018,Wellein_2009}.

Fig.~\ref{fig:mpi-synchronization} gives an overview of all necessary MPI
synchronization for the Euler update.
The synchronization of the vectors $\alpha_i$, $\bR_i$ and $\bU_i$ and the
matrix $l_{ij}$ over MPI ranks incurs point-to-point communication and forces an
individual MPI rank to wait until all necessary data has arrived. In
addition, the computation of the maximal admissible step size,
$\tau_{\text{max}}$, requires an \emph{MPI Allreduce} operation and thus
incurs an MPI barrier after step 2 during which all MPI ranks have to wait.
While the MPI
barrier for computing $\tau_{\text{max}}$ is unavoidable, it is possible to
mitigate the synchronization overhead to a certain degree by scheduling the
synchronization of vectors and matrices as soon as possible. We refer to
Section~\ref{subse:communication_hiding} for a detailed discussion how this
can be achieved is in our approach.
Benchmark results for weak and strong scaling are given in
Section~\ref{sec:benchmarks}.

\begin{remark}
  \label{rem:overlap}
  An additional measure to reduce the number of MPI synchronizations is to
  increase the overlap of shared cells between neighboring MPI ranks. This
  would allow to remove most of the synchronization steps outlined in
  Fig.~\ref{fig:mpi-synchronization} with the exception of the (essential) MPI
  barrier after step 2 that is necessary to determine $\tau_{\max}$. We do
  not pursue this optimization in the present work because it increases the
  amount of computations, the limiting resource away from the strong scaling
  limit. Our benchmarks in Sec.~\ref{sec:benchmarks} show that the MPI
  synchronization overhead is small, such that the choice does not pose a real
  limitation.
\end{remark}

\subsection{Instruction-level SIMD vectorization}
\label{subse:vectorization}

In order to exploit the SIMD capabilities offered by modern CPUs reliably
and to an appreciable degree also for more complex algorithms
and data dependencies one is usually forced to
``\emph{vectorize by hand}'' \cite{Kronbichler_2019} instead of relying on
the auto-vectorization capabilities of optimizing compilers. This
can be achieved in a portable manner by exploiting the C++ class mechanism
and operator overloading. We refer the reader to \cite{dealii91,
dealiicanonical,Kronbichler_2012} for details on the implementation of
deal.II's \texttt{VectorizedArray} class template that provides such a
facility.
\footnote{%
  The \texttt{VectorizedArray} class is conceptually
  very similar to the \texttt{std::simd} class that is currently
  considered for inclusion into the upcoming \texttt{C++23} standard; see
  \cite{Hoberock_2019}.
}

The first design decision that we have to make when expressing
Alg.~\ref{alg:euler} in vectorized form is to decide which part of the computation can
be meaningfully fused together. Here, we have multiple options. We could,
for example, decide to introduce parallel SIMD instructions within the
innermost loop, or to parallelize over the loop index $j$, viz.,\\
\begin{algorithm}[H]
  \begin{minipage}{0.48\textwidth}
    \DontPrintSemicolon
    \For{$i=1$, \ldots, $\NN$}{
      \For{$j\in\Ii$}{
        \tcp*[l]{SIMD instructions parallelizing:}
        \emph{Computation involving index $i$ and $j$.}
      }
    }
  \end{minipage}%
  \begin{minipage}{0.48\textwidth}
    \DontPrintSemicolon
    \For{$i=1$, \ldots, $\NN$}{
      \tcp*[l]{SIMD instructions fusing the for loop:}
      \For{$(j,j+1,\ldots,j+k)\in\Ii$}{
        \emph{Comp. involving index $i$ and $(j,j+1,\ldots,j+k)$.}
      }
    }
  \end{minipage}
\end{algorithm}
\noindent However, these two approaches have the significant drawback that
they would require carefully handwritten assembly to achieve good
utilization of vector registers. The difficulties are caused by complex
data dependencies and because the number of indices in $\Ii$ or the number
of equations $d+2$ might not be divisible by the width $k$ of the SIMD
registers. We opt for a different strategy by applying SIMD to the outer
loop over $i$:\\
\begin{algorithm}[H]
  \DontPrintSemicolon
  \tcp*[l]{SIMD instructions fusing the for loop:}
  \For{$(i,i+1,\ldots,i+k) \in [1,\NN]$}{
    \For{$(j_{1},j_{2},\ldots,j_{k})\in\Ii\times\mathcal{I}(i+1)\times\ldots\times\mathcal{I}(i+k)$}{
      \emph{Computation involving indices $(i,i+1,\ldots,i+k)$ and
        $(j_1,j_2,\ldots,j_k)$.}
    }
  }
\end{algorithm}
The main advantage of this scheme is that the operations on several points in
the stencil are more uniform, leading to a good utilization of vector
units. The idea to apply vectorization at an outer loop with additional
similarity is conceptually similar to vectorization across elements popular of
matrix-free methods~\cite{Kronbichler_2012,Kronbichler_2019,Sun_2020}.
This approach has the minor drawback to require the set $\Ii$
to be of equal size for all indices that are processed at the same time, and
that the limiter involving the quadratic Newton iteration has to be adapted to
process multiple states at the same time. We point out that this can be
achieved with relatively minor modifications to the (mathematical)
algorithms presented in Section~\ref{sec:scheme}. For example,
Alg.~\ref{alg:limiter} contains a number of \emph{ternary} operations of
the form
\begin{center}
  \texttt{if (condition), select A, otherwise select B},
\end{center}
which can be efficiently implemented with SIMD masking techniques
\cite{Hoberock_2019}
\footnote{%
  Convenience functions implementing ternary operations on SIMD vectorized
  data are readily available in deal.II via function wrappers such as
  \texttt{compare\_and\_apply\_mask<SIMDComparison::less\_than>(a,\,b,\,c,\,d)}
  which is equivalent to \texttt{(a<b)\,?\,c\,:\,d}. These ternary
  operations are expected to eventually become ``first-class citizens'' in
  a future C++23 standard with the introduction of \texttt{std::simd} and
  corresponding \texttt{operator?:} overloads.
}.
Branching on data in the algorithm only occurs with the break statements in
the for loop in Alg.~\ref{alg:limiter}. These have to be modified to check
whether the condition is simultaneously fulfilled for all states of the
SIMD vector. This implies that some of the states, which the limiter works
on in parallel, might undergo an additional Newton iteration in the algorithm
despite convergence.

Another point to consider is the fact that parallelizing over the outer
loop comes at the cost of increased pressure on caches which will be
discussed in more detail in Section~\ref{sec:benchmarks}.

\subsection{Local indexing of degrees of freedom and a SIMD optimized sparsity pattern}
\label{subse:local_index}
\begin{figure}
  \begin{tikzpicture}[scale=0.8]
    \fill[fill=blue, fill opacity=0.1] (0.0,0.5) rectangle (1.5,-1.5);
    \fill[fill=blue, fill opacity=0.1] (7.0,0.5) rectangle (8.5,-1.5);
    \fill[fill=red, fill opacity=0.1] (8.5,0.5) rectangle (10.,-1.5);
    \node at (0.75, -1.25) {export};
    \node at (7.75, -1.25) {export};
    \node at (9.25, -1.25) {import};
    \draw[thick, {Bracket[width=4mm,line width=1pt,length=1.0mm]}-{Parenthesis[width=4mm,line width=1pt,length=1.5mm]}] (0, 0) -- (10, 0);
    \node at ( 10.0, -0.6) {$\NN_{\,lr}$};
    \draw[thick, {Bracket[width=4mm,line width=1pt,length=1.0mm]}-{Parenthesis[width=4mm,line width=1pt,length=1.5mm]}] (0, 0) -- (8.5, 0);
    \node at ( 8.5, -0.6) {$\NN_{\,lo}$};
    \draw[thick, {Bracket[width=4mm,line width=1pt,length=1.0mm]}-{Parenthesis[width=4mm,line width=1pt,length=1.5mm]}] (0, 0) -- (7, 0);
    \node at ( 7.0, -0.6) {$\NN_{\,i}$};
    \draw[thick, {Bracket[width=4mm,line width=1pt,length=1.0mm]}-{Parenthesis[width=4mm,line width=1pt,length=1.5mm]}] (0, 0) -- (1.5, 0);
    \node at ( 0.0, -0.6) {$0$};
    \node at ( 1.5, -0.6) {$\NN_{\,e}$};
  \end{tikzpicture}
  \caption{%
    Local index handling: On each MPI rank we enumerate all locally
    relevant dofs with a local index $[0,\NN_{\text{\,lr}})$ subject to the
    following contraints: $\NN_{\text{\,i}}$ is a multiple of $k$, the
    width of the SIMD registers and the interval $[0,\NN_{\text{\,i}})$
    only contains dofs with standard connectivity ($\#(\Ii)=3$, $9$, or
    $27$). A subsequent renumbering ensures that $[0,\NN_{\text{\,e}})$
    contains all exported degrees of freedom of the internal range. Ranges
    with dofs that have to be exported or imported during MPI synchronization
    are marked in blue and red, respectively.}
  \label{fig:local_index_numbering}
\end{figure}
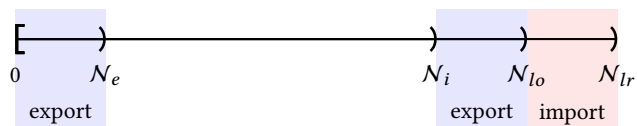

A common strategy for handling a global numbering of degrees of freedom is
to assign a contiguous interval of \emph{locally owned} dofs to an
individual MPI rank in a 1:1 fashion, and a typically larger set of
\emph{locally relevant} dofs described by the access pattern of the owned
rows, $\big\{j\in\Ii\;:\;i\text{ is a locally owned
dof}\big\}$~\cite{dealiicanonical}. The latter index set includes the
foreign dofs, also called ghost dofs, necessary to update the locally owned range on the
respective MPI rank.

This global numbering is then transformed into a numbering of dofs local to
each MPI rank. It starts at 0 so that the index can be directly used as an
offset into the the underlying storage in memory. In the following we adopt
the convention that the local numbering range is comprised of two disjunct
intervals: $[0,\NN_{\text{\,lo}})$ contains all locally owned dofs and
$[\NN_{\text{\,lo}}, \NN_{\text{\,lr}})$ contains all locally relevant dofs
that are not locally owned.

The SIMD parallelization approach outlined in the previous section requires
a uniform stencil size, i.\,e., $\#(\Ii)=\text{const.}$, over the region of
indices that will be vectorized. We ensure this property by applying a
local renumbering of the locally-owned index range
$[0,\NN_{\text{\,lo}})$ as follows. We sort the interval into a range
$[0,\NN_{\,i})$ of \emph{internal} degrees of freedom with standard
connectivity that we
characterize by $\#(\Ii)=3$, $9$, or $27$, depending on dimension.
Correspondingly, the interval $[\NN_{\,i},\NN_{\text{\,lo}})$ contains dofs
that have a different stencil size. We round $\NN_{\text{\,i}}$ down to the
next integral multiple of $k$, the width of the SIMD registers, and schedule
the loop with full SIMD width. As a final
step the interval $[0,\NN_{\text{\,i}})$ is further rearranged so that
$[0,\NN_{\text{\,e}})$ contains all \emph{exported} dofs within the
internal number range, that is, all internal dofs that are also part of a
foreign MPI rank's locally relevant index range and thus have to be
exchanged during MPI synchronization. A graphical summary is given in
Fig.~\ref{fig:local_index_numbering}.
\begin{figure}
  \begin{tikzpicture}
    \node[align=center] at (5.0, 1.75)
      {$\underline{A}_{ij}\;=\;\big(\textcolor{red}{A_{ij}^1},\,
        \textcolor{blue}{A_{ij}^2},\,\textcolor{green}{A_{ij}^3}\big)^T$};
    \fill[fill=blue, fill opacity=0.1] (-6.3, 1.95) rectangle (2.5,-0.1);
    \node[align=center] at (-4.5, 0.90)
      {SIMD parallel region\\[0.2em]SELL-C $\sigma$ format\\[0.2em](fixed row size)};
    \draw[->, thick] (-6.4,  0.9) to [bend right=30] (-7.5, -3.3);
    \fill[fill=red, fill opacity=0.1] (-2.5,-0.3) rectangle (6.3,-1.95);
    \node[align=center] at ( 4.5, -1.1)
      {non-vectorized region\\[0.2em]CSR format\\[0.2em](variable row size)};
    \draw[->, thick] (6.4, -1.1) to [bend left=20] (7.0, -4.3);
    \node at (0.0,-2.5) {$\underline{A}_{ij}$ stencil};
    \node at (0.0, 0.0) {$%
      \mleft[\begin{array}{ccccc|cccc}
        x & 1 & 2 & 3 & 4 & &   &   &  \\
          & x & 1 & 2 & 3 & &   & 4 &  \\
        1 &   & x &   & 2 & &   & 3 & 4\\
        1 & 2 & 3 & x &   & & 4 &   &  \\
          & & & & \ddots & & & &  \\
          \hline
          & & & & & \ddots & & &  \\
          & 1 & 2 &   & 3 & & x & 4 & 5\\
          &   &   &   & 1 & & 2 & x &  \\
          &   &   & 1 & 2 & & 3 & 4 & x\\
      \end{array}\mright]$};
    \node at (-3.0, 2.00) {$0$};
    \node at (-3.0,-0.20) {$\NN_{\text{\,i}}$};
    \node at (-3.0,-2.00) {$\NN_{\text{\,lr}}$};
    \node at (-2.65,2.40) {$0$};
    \node at ( 0.25,2.40) {$\NN_{\text{\,i}}$};
    \node at ( 2.65,2.40) {$\NN_{\text{\,lr}}$};
    \node at (-2.0,-3.5) {$%
      \underbracket{\textcolor{red}{xxxx}\,\textcolor{blue}{xxxx}\,\textcolor{green}{xxxx}}\,
      \underbracket{\textcolor{red}{1111}\,\textcolor{blue}{1111}\,\textcolor{green}{1111}}\,
      \underbracket{\textcolor{red}{2222}\,\textcolor{blue}{2222}\,\textcolor{green}{2222}}\,
      \underbracket{\textcolor{red}{3333}\,\textcolor{blue}{3333}\,\textcolor{green}{3333}}\,
      \underbracket{\textcolor{red}{4444}\,\textcolor{blue}{4444}\,\textcolor{green}{4444}}
      \;\ldots
    $};
    \node at (2.0,-4.5) {$%
      \ldots\;
      \underbracket{\textcolor{red}{x}\,\textcolor{blue}{x}\,\textcolor{green}{x}}\,
      \underbracket{\textcolor{red}{1}\,\textcolor{blue}{1}\,\textcolor{green}{1}}\,
      \underbracket{\textcolor{red}{2}\,\textcolor{blue}{2}\,\textcolor{green}{2}}\,
      \underbracket{\textcolor{red}{3}\,\textcolor{blue}{3}\,\textcolor{green}{3}}\,
      \underbracket{\textcolor{red}{4}\,\textcolor{blue}{4}\,\textcolor{green}{4}}\,
      \underbracket{\textcolor{red}{5}\,\textcolor{blue}{5}\,\textcolor{green}{5}}\,
      \underbracket{\textcolor{red}{x}\,\textcolor{blue}{x}\,\textcolor{green}{x}}\,
      \underbracket{\textcolor{red}{1}\,\textcolor{blue}{1}\,\textcolor{green}{1}}\,
      \underbracket{\textcolor{red}{2}\,\textcolor{blue}{2}\,\textcolor{green}{2}}\,
      \underbracket{\textcolor{red}{x}\,\textcolor{blue}{x}\,\textcolor{green}{x}}\,
      \underbracket{\textcolor{red}{1}\,\textcolor{blue}{1}\,\textcolor{green}{1}}\,
      \underbracket{\textcolor{red}{2}\,\textcolor{blue}{2}\,\textcolor{green}{2}}\,
      \underbracket{\textcolor{red}{3}\,\textcolor{blue}{3}\,\textcolor{green}{3}}\,
      \underbracket{\textcolor{red}{4}\,\textcolor{blue}{4}\,\textcolor{green}{4}}
    $};
    \draw[-{Latex}, thick] (-2.4, -5.3) -- (-2.4, -4.8);
    \node at (-1.7,-5.3) {rind ptr};
    \draw[-{Latex}, thick] (1.6, -5.3) -- (1.6, -4.8);
    \draw[-{Latex}, thick] (3.625, -5.3) -- (3.625, -4.8);
  \end{tikzpicture}
  \caption{%
    A SIMD optimized sparsity pattern visualized for (the hypothetical case
    of) a standard connectivity of $\#(\Ii)=5$ and a width $k=4$ of the
    SIMD registers. The SIMD vectorized index range $[0,\NN_{\text{\,i}})$
    is stored in sliced ELL format as an ``array of struct of array'' as follows:
    at the innermost 'array' level, we group the same entry from $k$
    consecutive rows together; next come the different components in case
    we have a multi-component matrix, i.\,e., the ``struct'' level
    groups the components next to the inner array of row data; finally,
    Finally, the outer array arranges the different components in an ELL
    storage format. The non-vectorized region is stored in a CSR storage
    format (i.e., SELL-1) on the outer layer grouping the same struct level (that
    organizes the components of a multi-component matrix together).
  }
\end{figure}
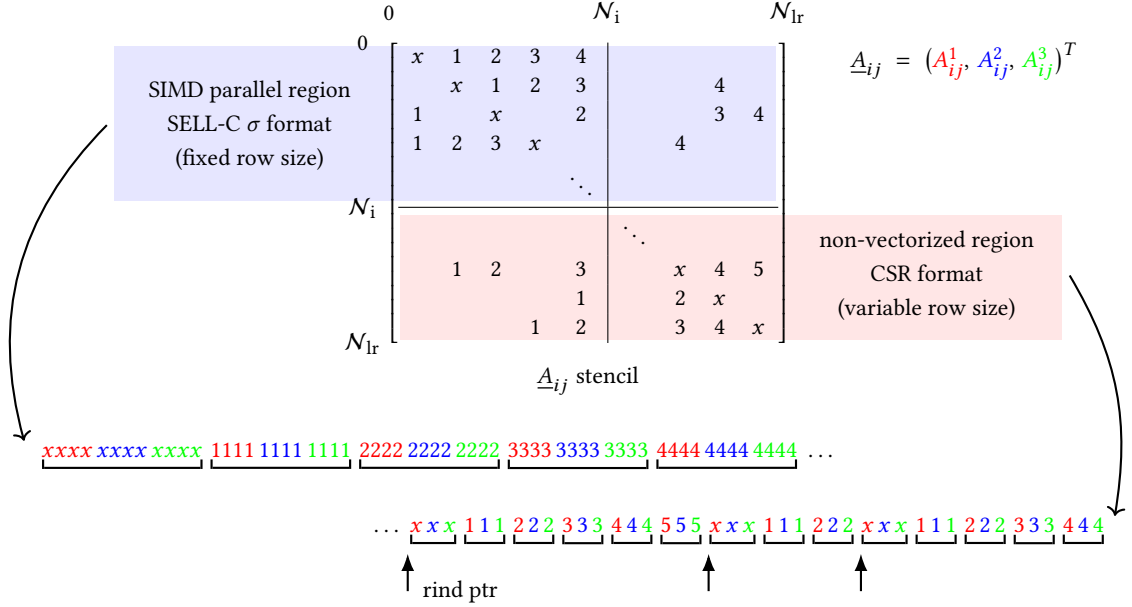

\begin{remark}
It would be possible to also vectorize the remainder loop
$[\NN_{\,i},\NN_{\text{\,lo}})$, for
example by using an elaborate \emph{masking} strategy, or a fill with dummy
values to account for differing stencil sizes. The latter comes with the
additional challenge that a suitable neutral element for all operations involved
in the nonlinear stencil update would be needed. Thus, we opt for the more pragmatic
solution of not vectorizing the remainder. We justify this approach with two
observations. First of all, the number of affected degrees of freedom is
asymptotically small, typically less than 3\% of all degrees of freedom for
moderately sized problems (see Section~\ref{sec:benchmarks}). Secondly,
treating boundary dofs separately allows for some further optimization in
Alg.~\ref{alg:euler}. For example, the symmetrization of the wavespeed estimate
coming from the Riemann solver in step 1 can be skipped entirely
\cite{gnpt_second_order_2018}.
\end{remark}

Based on our vectorization approach we propose an optimized sparsity
pattern that ensures a linear traversal through the storage region of all
matrices in memory. The sparsity pattern handles vector-valued matrix
entries as needed for the $\bc_{ij}$ matrix: The SIMD-vectorized index
range $[0,\NN_{\text{\,i}})$ is stored in sliced-ELL format~\cite{Kreutzer_2014}
as an ``\emph{array of
struct of array}'' as follows: at the innermost 'array' level, we group the
same entry from $k$ consecutive rows together; next come the different
components in case we have a multi-component matrix, i.\,e., the
``\emph{struct}'' level groups the components next to the inner array of
row data; finally, the outer array arranges the different
components in an ELL storage format. The non-vectorized region is stored in
a CSR storage format on the outer layer grouping the same struct level
(that organizes the components of a multi-component matrix toegether).

The proposed storage scheme is a variant of the SELL-C-$\sigma$ sparsity
pattern proposed by \citet{Kreutzer_2014}. This format is well-suited for both
contemporary CPU and GPU architectures with appropriate values for the
parameter $C$ of the inner length of slices, see also the recent analysis of
\citet{Anzt_2020}. As indicated above, the slice length proposed in this work
corresponds to the widest SIMD register in doubles, e.g., 8 for AVX-512. This
ensures that vector loads can be performed for all matrix entries. The
classification of the rows corresponds to a large window $\sigma$ for the row
lengths in the CELL-C-$\sigma$ format spanning all locally owned degrees of
freedom. Thus, the fill in the sliced ELL region is always optimal. However,
we switch to slice length $C=1$ in the irregular rows for the present
contribution, given their small share on the overall rows and the reasonable
performance of scalar operations on general-purpose CPU architectures
considered here.

\subsection{Storage of state vectors}
\label{subse:statevecs}

On each node of the computational domain, the state vectors $\bUni$ as well
as the temporary vector $\bR_i$ contain $d+2$ components. The two storage
options are (i) a struct-of-array, keeping $d+2$ separate vectors for each
component, or (ii) an array-of-struct, a single vector which puts the $d+2$
components of a single node adjacent in memory. We propose the
array-of-struct storage option for the following reasons:
\begin{itemize}
  \item The data exchange routines of conventional MPI-parallel vectors
    straight-forwardly combine the data from all components into the same
    point-to-point messages, without manually collecting the data before
    sending. This slightly reduces latency in the strong scaling limit, see
    also the discussion in Fischer et al.~\cite{Fischer_2020}.
  \item The vectorized data access due to contiguous indices $i$ in the
    struct-of-array variant would only help the access to row data in the
    outer $i$ loops, whereas the more frequent column access in the inner $j$
    loops would still appear as indirect gather access unless the mesh is
    completely structured. Thus, the
    array-of-struct format leads to more contiguous access for unstructured
    meshes. This reduces pressure on the translation-lookaside buffer (TLB)
    and increases hardware prefetching efficiency considerably.
  \item The necessary transpose operations from the stored array-of-struct to
    the SIMD struct-of-array format of multiple row data can be done with two
    shuffle-type instructions per entry for chunks of four double-precision
    values.
\end{itemize}
Benchmarks of the code with the two variants revealed that the chosen
struct-of-array storage makes the evaluation considerably faster. For example
for the access to $\bUnj$ in step 1 of Alg.~\ref{alg:euler} computed with 28.6
million $\mathcal Q_1$ mesh points followed over 1302 Euler step evaluations
on 80 cores, the run time is reduced from 599 seconds to 391 seconds, all
other parts equal.

\subsection{MPI communication hiding}
\label{subse:communication_hiding}

A single explicit Euler update (Alg.~\ref{alg:euler}) requires a number of
MPI synchronization events between individual steps of the algorithm that
cannot continue until all foreign data of the locally relevant index range
is exchanged; see Fig.~\ref{fig:mpi-synchronization}. In order to minimize
latency incurred by the MPI synchronization we use a common \emph{MPI
communication hiding} \cite{Brightwell_2005} technique: The non-SIMD vectorized part
$[\NN_{\,i},\NN_{\text{\,lo}})$ and the vectorized subregion $[0,\NN_{\,e})$
are computed first which allows to start an asynchronous MPI
synchronization process early. The computation can then continue with
computing the large vectorized index region $[\NN_{\,e},\NN_{\text{\,i}})$
while the MPI implementation exchanges messages. We use a simple thread
synchronization technique centered around a \texttt{std::atomic} for the
actual implementation in context of our hybrid thread-process
parallelization, see Alg.~\ref{alg:synchronization}.\\
\begin{algorithm}[tb]
  \DontPrintSemicolon
  \SetKwBlock{Parallel}{thread parallel region}{end}
  \SetKwFor{ParallelFor}{parallel for}{do}{end}
  \textbf{std::atomic<unsigned int>} n\_threads\_ready $\;\leftarrow\;$ 0\;
  \Parallel{
    \ParallelFor{$i\,\in[\NN_{\,i},\,\NN_{\,lo})$}{
      \tcp*[l]{Compute serial part.}
    }
    \textbf{bool} this\_thread\_ready $\;\leftarrow\;$ false\;
    \ParallelFor{$i\,\in[0,\,\NN_{\,i})$}{
      \If{\emph{[unlikely]} (\,this\_thread\_ready == false\,) and
      (\,$i\ge\NN_{\,e}$\,)}{
        this\_thread\_ready $\;\leftarrow\;$ true\;
        \If{++n\_threads\_ready == n\_threads}{
          \tcp*[l]{Initialize MPI synchronization.}
        }
      }
      \tcp*[l]{compute SIMD vectorized part.}
    }
  }
  \tcp*[l]{Wait for MPI synchronization to finish.}
  \caption{MPI communication hiding in thread-parallel context. A
    thread-local boolean \texttt{this\_thread\_ready} is used to avoid
    unnecessary thread-synchronization and ensures that the if condition in
    the second parallel for loop is only entered exactly once on every
    thread. The default memory model of \texttt{std::atomic} then ensures
    that the condition \texttt{n\_threads\_ready == n\_threads} is true on
    exactly one thread.}
  \label{alg:synchronization}
\end{algorithm}

\subsection{Vectorized power function}
\label{subse:vectorized_pow}

The nonlinear update step shown in Alg.~\ref{alg:euler} makes heavy use of
transcendental \texttt{pow()} operations when computing the entropy-viscosity
commutator described in Sec.~\ref{subsec:entropy-viscosity-commutator} and in
the limiter described in Sec.~\ref{subsec:specific-entropy-limiter}. Such
transcendental operations are computationally expensive \cite{Fog_2020}. As
detailed in Sec.~\ref{subse:roofline} below, an update step consists of about
4--8 \texttt{pow()} invocations per non-zero entry in the stencil (nnz). It is
thus of paramount importance to use an optimized and vectorized \texttt{pow()}
implementation. In our benchmark code we choose the C++ Vector Class Library%
\footnote{\url{https://github.com/vectorclass}}
by Fog et al.~\cite{Fog_2020}.

In order to assess the computational properties, we ran a microbenchmark that
repeatedly calls \texttt{pow(x,1.4)} over a vector of 20,480 random numbers
between 1 and 2. The reciprocal throughput per entry is for the naive
(non-vectorized) implementation using the standard library implementation
\texttt{std::pow}%
\footnote{\url{https://gcc.gnu.org/onlinedocs/libstdc++/}}
gives an execution time of 73 nanoseconds at a clock frequency of 2.8 GHz.
The vectorized version of the Vector Class Library achieves a reciprocal
throughput of 8.1\,ns at a clock frequency of 2.0 GHz (the maximum frequency
for AVX-512 heavy code when loading all cores of an Intel Cascade Lake machine
according to Table~\ref{tab:systems}) or 65\,ns (130 clock cycles) per call.
The recorded throughput is relatively close to more heavily optimized code for
\emph{multiple} \texttt{pow()} invocations with the Intel\textregistered Math
Kernel Library (mkl)
\footnote{\url{https://software.intel.com/content/www/us/en/develop/tools/math-kernel-library.html}}
of about 4.4\,ns, 4.3\,ns, and 2.1\,ns (for ``high accuracy'', ``low
accuracy'', and ``enhanced performance'' variants).  We suspect that the
performance for the mkl library is higher due to significantly better
pipelining of instructions for consecutive \texttt{pow()} operations. In order
to realize this throughput in Alg.~\ref{alg:euler}, a substantial rewrite of
the algorithm (such that \texttt{pow()} operations of multiple columns are
executed in succession) would be necessary, a task we leave for future
research and modifications discussed in the outlook in
Sec.~\ref{sec:conclusion}.


\section{Benchmarks and results}
\label{sec:benchmarks}
\begin{figure}
  \centering
  \includegraphics[width=0.7\textwidth]{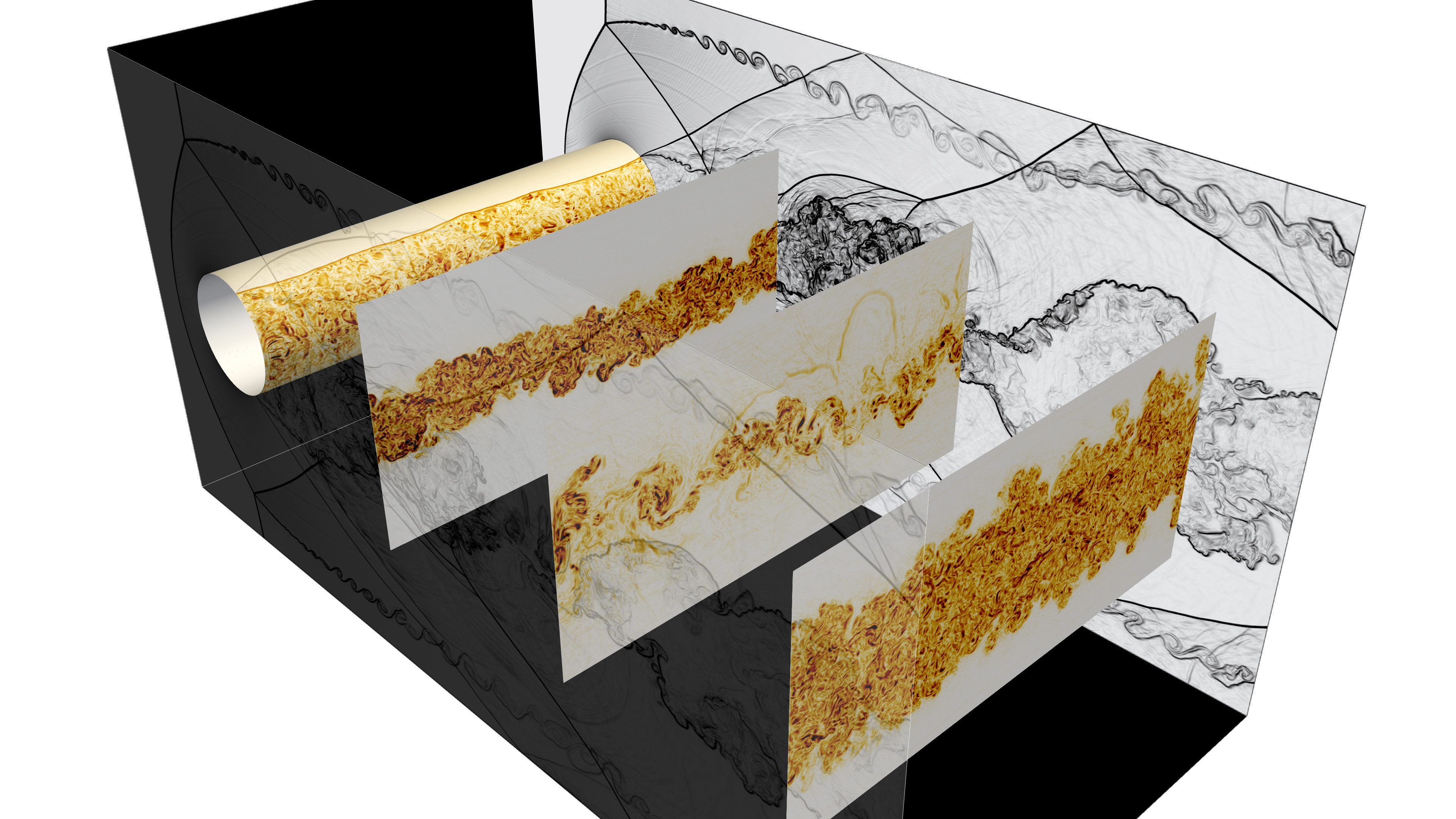}
  \caption{Temporal snapshot of a longer computation of a 3D Mach 3
    compressible Euler flow around a disc discretized with 1.8\,B Qdofs at
    $t=5.0$. The two vertical outer cutplanes show a \emph{Schlieren plot},
    i.\,e., the magnitude of the gradient of the density is shown on an
    exponential scale from white (low) to black (high). All other cutplanes
    show the magnitude of the vorticity on a white (low) - yellow (medium)
    - red (high) scale. The computation was done with an earlier, not fully
    optimized version of the solver and ran on 30720 MPI ranks with an
    average time-step size of 6.0e-05. The code achieved an average
    throughput of 969 QDofs per second (0.04M gridpoints per second per
    CPU) with a second-order SSP Runge-Kutta time integrator, in contrast to
    the third-order variant suggested in this paper (see
    Sec.~\ref{subsec:ssp-rk}).}
  \label{fig:cyl2}
\end{figure}

All computations are performed for a 3D benchmark configuration
\cite{gnpt_second_order_2018}, similar to the 2D configuration shown in
Fig.~\ref{fig:cyl}, that consists of a supersonic (air) flow at Mach 3 in a
rectangular parallelepiped of size $[0,4]\times[-1,1]\times[-1,1]$ past a
cylinder with radius $0.25$ is centered along $(0.6, 0, z)$, $z\in[-1,1]$.
The computational domain is meshed with an unstructured hexahedral coarse
mesh and trilinear $\mathcal Q_1$ elements consisting of 208 gridpoints, or
nodal degrees of freedom (Qdofs). A higher resolution is obtained by
subdividing every hexahedron into 8 children an appropriate number of
times, using a cylindrical manifold to attach newly generated nodes along
the cylinder to the curved surface. Fig~\ref{fig:cyl2} shows a temporal
snapshot at time $t=5.0$ of a typical computation with 1.8B Qdofs.

The hardware used for the experiments in this section is described in
Table~\ref{tab:systems}. Both machines are deployed in the form of compute nodes
with dual-socket configurations (two CPUs per compute node) with a high-speed
network interconnect (Infiniband/Omnipath). The Intel Cascade Lake system
has a machine balance of 14.2 Flop/Byte computed from the peak arithemtic
throughput and the STREAM triad bandwidth%
\footnote{\url{https://www.cs.virginia.edu/stream/ref.html}}
compared to 17.2 Flop/Byte on the Intel Skylake system.%
\footnote{%
Note that for the Intel Cascade Lake system, the gap between the
theoretical memory bandwidth and the actually measured STREAM bandwidth is
higher due to the particular hardware configuration (single-rank vs
dual-rank memory modules).
}
\begin{table}
  \caption{Hardware used for the computational experiments and benchmarks.
    The STREAM triad bandwidth is measured with streaming stores, i.e., it
    reports the actually transferred data between the cores and the
    memory.}
  \label{tab:systems}
  \begin{tabular}{lll}
    \toprule
     & \textbf{Intel Cascade Lake} & \textbf{Intel Skylake}\\
    \midrule
    Model name & Xeon Gold 6230 & Xeon Platinum 8174\\
    Cores / compute node & $2\times 20$ & $2\times 24$ \\
    SIMD width  & 512 bit (AVX-512) & 512 bit (AVX-512)\\
    Turbo mode & enabled & disabled \\
    Clock frequency scalar & 2.8 GHz & 2.3 GHz\\
    Clock frequency AVX-512 & 2.0 GHz & 2.3 GHz\\
    L2 + L3 cache / core & 1 MiB + 1.375 MiB & 1 MiB + 1.375 MiB\\
    Arithmetic peak with AVX-512 / compute node & 2,560 GFlop/s & 3,532 GFlop/s\\
    Peak memory bandwidth / compute node & 282 GB/s & 256 GB/s\\
    STREAM triad bandwidth from RAM / compute node & 180 GB/s & 205 GB/s\\
    \bottomrule
  \end{tabular}
\end{table}

\subsection{Roofline performance prediction and kernel selection}
\label{subse:roofline}

The mathematical description of Alg.~\ref{alg:euler} allows some freedom in
rearranging computations between individual loops. In order to find the
algorithm variant with the best performance, we need to identify the
limiting computational resource. A stencil code such as the one presented
in Alg.~\ref{alg:euler} of sufficient local size, i.\,e., with more than a
few thousand Qdofs per MPI rank, is operated in the throughput regime with
respect to communication between the compute nodes. The two primary bottlenecks are thus
data access, which is governed by the bandwidth from main memory or caches,
and the in-core execution, which can be represented by the roofline
performance model~\cite{Williams_2009}.

\subsubsection{Data access}

In Table~\ref{tab:memory} we list the expected memory access of the stages
in the final optimized version of the algorithm. All numbers are given as
reads and writes per per non-zero entry in the stencil (nnz). The predicted
access is reported separately for read transfer (labeled `r' in the table),
writes (labeled `w' in the table), and the \emph{read-for-ownership
transfer} \cite{Hager_2011}, labeled `rfo' in the table. The
read-for-ownership transfer adds additional read transfer for data that is
only written. We use non-temporal (streaming) stores for the matrices
$d_{ij}$ of step 1, $P_{ij}$ in step 4 and $l_{ij}$ in steps 4 and 5 to
avoid the read-for-ownership transfer, but regular stores for the vector
data $\bU^{n+1}$ and $\bR^{n+1}$. The performance prediction is based on the
following assumptions:
\begin{itemize}
  \item
    all big data structures need to be fetched from RAM memory in their
    entirety for every evaluation step; this includes the matrices $m_{ij},
    \beta_{ij}, \mathbf{c}_{ij}$ and the underlying sparsity pattern as
    well as the global vectors $\bU^n$, $\bR^n$, $\bU^{n+1}$, and the
    vector for the lumped mass matrix;%
    \footnote{%
      This assumption is justified because the loops are not overlapped and
      the size is big enough to exceed caches by at least a factor of 10.
    }
  \item
    access to column data of $\bUnj$ and $\bR^{n}_j$, the inverse mass
    matrix and $\alpha_j$ exhibits perfect caching;
  \item
    access to the transposed matrix entries $d_{ji}$ and $l_{ji}$ in steps
    2, 5 and 6, respectively, exhibits perfect caching with perfect spatial
    locality.
\end{itemize}

The last two assumptions regarding data locality of column access are
similar to the layer conditions found in high-performance implementations
of finite difference stencils~\cite{Hager_2011}. For example, for the 2D
five-point stencil the layer criterion relates the spatial distance of an
entry $(i,j)$ to the grid neighbors $(i+1,j)$, $(i-1,j)$, $(i,j+1)$,
$(i,j-1)$ to the cache size. In order to only load one data item per
update, e.\,g., the $(i,j+1)$ entry during a lexicographic grid traversal,
the cache must be large enough to store two full rows of entries ($2 n_x$
items, where $n_x$ is the number of gridpoints in $x$-direction). For
larger mesh sizes the loop must be tiled. The main difference to the
present (finite-element) algorithms is the fact that they are written for
unstructured meshes with indirect addressing of column data. Thus, a
corresponding 3D layer condition for a structured grid requiring that $2
n_x n_y$ items fit into cache has to be modified. A simple imitation of
lexicographic numbering for unstructured meshes is obtained by a
Cuthill-McKee ordering of the unknowns \cite{Cuthill_1969}. We can assume
that the Cuthill-McKee reordering maintains a bandwidth of approximately
$n^{2/3}_\text{local}$ unknowns per row, where $n_\text{local}$ is the
number of DoFs per MPI rank. A modified line criterion could thus be the
requirement to hold $2\,n_\text{local}^{2/3}$ entries in cache. This
implies for the example presented in Table~\ref{tab:memory} with an average
local size of $n_\text{local}=358,208$ dofs that about 10,200 entries have
to be kept in cache. A state vector $\bU^n$ holds five variables per entry.
With  8 bytes per double this equates to 400 kiB. Given that the
architecture in use provides around 2.4 MiB of L2 and L3 cache combined, we
can expect that the modified line criterion is mostly fulfilled in step 1
of the algorithm. On the other hand, in step 4, both vectors $\bU^n$ and
$\bR^n$ amounting to 800 kiB according to the modified layer criterion are
required to be maintained in cache, in addition to streaming through the
matrices $d_{ij}^{L,n}$ and $m_{ij}$ at the same time. Realistically, step
4 will involve some additional transfer from main memory due to cache
eviction.

\begin{table}
  \caption{Expected memory transfer and measured performance on a
    simulation with 29m nodes over 434 time steps (1302 RK stage
    evaluations), run on 80 Intel Cascade Lake cores (2 compute nodes). Memory
    bandwidth for STREAM triad is 360 GB/s.}
  \label{tab:memory}
  \begin{tabular}{lccccc}
    \toprule
    & & \multicolumn{3}{c}{\textbf{measurement with likwid}} & \textbf{prediction}\\
    \cmidrule(lr){3-5} \cmidrule(lr){6-6}
    & time  & bandw & read / write & r / w barrier & read / write\\
    & [s] & [GB/s] & [double / nnz] & [double / nnz] & [double / nnz]\\[0.3em]
    step 0: entropies & 9.65  & 239 & 0.22r + 0.07w & 0.29r + 0.09w &  0.19r + 0.07w + 0.07rfo\\
    step 1: offdiagonal $d^L_{ij}$, $\alpha_i$ & 391.4 & 132 & 5.83r + 0.74w & 5.46r + 0.65w & 4.72r + 0.56w + 0.04rfo\\
    step 2: diagonal $d^L_{ii}$, $\tau_n$ & 62.0 & 304 & 1.95r + 0.46w & 2.44r + 0.56w & 1.74r + 0.48w\\
    step 3: low-order update & 277.0 & 222 & 7.24r + 0.53w & 7.21r + 0.51w & 5.87r + 0.48w + 0.48rfo\\
    step 4: $P_{ij}$, $l_{ij}$ & 317.8 & 248 & 4.43r + 6.03w & 4.31r + 6.04w & 3.24r + 6.00w\\
    step 5: h.-o. update, next $l_{ij}$ & 268.7 & 260 & 8.06r + 1.20w & 8.21r + 1.21w & 6.80r + 1.19w\\
    step 6: final high-order update & 132.5 & 394 & 6.71r + 0.21w & 7.98r + 0.21w & 6.69r + 0.19w\\
    \bottomrule
  \end{tabular}
\end{table}

Table~\ref{tab:memory} includes measurements of the memory read and write
access to the RAM memory, measured from hardware performance counters
recorded with the LIKWID tool~\cite{Treibig_2010}, version 5.0.1, using
an MPI-only experiment. The
numbers reported in the table are calculated from the absolute transfer
measured with LIKWID, divided by the number of time steps and stages per
time step and by the number of nonzero entries in the sparse matrix. The
result is further divided by 8, the number of bytes per double, to make the
numbers easily comparable to the transfer in terms of Alg.~\ref{alg:euler}.
The table includes two sets of measurements of markers around the
algorithmic part. The first part measures the sections as they appear in
the code. However, the numbers are inaccurate given a load imbalance of
5--15\% because the memory transfer is only recorded while the first core
of a 20-core CPU resides in the relevant section. If some of the other 19
cores take more time to complete the section (given the implicit barrier
via the MPI point-to-point communication at a later stage), the memory
transfer appears too low. This effect can be seen by the reads recorded for
step 6 of the algorithm, which should be close to step 5 in terms of the
transfer, but the reported number is 1.35 doubles less than the theoretical
number. In order to obtain more accurate data, we
performed a second experiment, labeled ``barrier'' in
Table~\ref{tab:memory}, where MPI barriers are placed around the
\texttt{LIKWID\_MARKER\_\{START/STOP\}} markers to ensure that only the
transfer of the respective section is measured. The write transfer, which
is of streaming character, is predicted very well. However, the actual read
transfer is by 15\%, 40\%, 14\%, 33\%, 21\%, and 19\% higher than the
best-case prediction for steps 1--6, respectively. For steps 1 and 3, the
excess transfer is contained because only a single vector $\bU^n$ and the
entropies, a total of 6 doubles per step, is accessed indirectly and one can
expect caches to mostly fit this access, with some minor deviations due to the
somewhat unstructured access in the Cuthill--McKee numbering and missing
spatial locality. For step 4, the access to both $\bU^n$
and $\bR^n$ leads to a larger deviation. In steps 2, 5, 6, the excess
transfer is due to the transpose access into a sparse matrix, where both
the limited size of the caches as well as the transfer of full cache lines
rather than single doubles are relevant.

\subsubsection{In-core execution}

The measured memory throughput in Table~\ref{tab:memory} demonstrates that
only step 6 is at the limit of the memory bandwidth of the architecture,
whereas all other steps are primarily limited by the execution inside the core. In
order to assess the arithmetic work done by the various stages,
Table~\ref{tab:flop} reports the main characteristics of the floating point
performance of the same computation. As discussed in
Section~\ref{subse:vectorized_pow}, the nonlinear update steps are heavy on
\texttt{pow()}, division and square root operations. Therefore, the
arithmetic peak performance of 4 Cascade Lake CPUs with 80 cores in total,
5,120 GFlop/s, is not attainable.

\begin{table}
  \caption{Main arithmetic components and measured performance on a
    simulation with 29m nodes over 434 time steps (1302 RK stage
    evaluations), run on 80 Intel Cascade Lake cores (2 compute nodes). Arithmetic peak
    is 5,120 GFlop/s.}
  \label{tab:flop}
  \begin{tabular}{lccccccc}
    \toprule
    & time & \multicolumn{4}{c}{\textbf{measurement with likwid}} & \multicolumn{2}{c}{\textbf{prediction}}\\
    \cmidrule(lr){3-6} \cmidrule(lr){7-8}
    & [s] & [GFlop/s] & [Flop/nnz] & [Flop/B] & IPC & [\texttt{pow()}/nnz] & [div/nnz]\\[0.3em]
    step 0: entropies & 9.65 & 848 & 8 & 2.6 & 1.32 & 0.07 & 0.04\\
    step 1: offdiagonal $d^L_{ij}$, $\alpha_i$ & 391.4 & 681 & 262 & 5.5 & 0.95 & 1.08 & 8.88 \\
    step 2: diagonal $d^L_{ii}$, $\tau_n$ & 62.0 & 17 & 1 & 0.04 & 1.65 & 0 & 0.04\\
    step 3: low-order update & 277.0 & 892 & 248 & 4.0 & 1.28 & 1 & 3.15\\
    step 4: $P_{ij}$, $l_{ij}$ & 317.8 & 571 & 183 & 2.2 & 1.16 & 1--2 (Newton) & 2--8\\
    step 5: h.-o. update, next $l_{ij}$ & 268.7 & 568 & 155 & 2.0 & 0.97 & 1--2 (Newton) & 2--8\\
    step 6: final high-order update & 132.5 & 91 & 12 & 0.18 & 0.17 & 0 & 0\\
    \bottomrule
  \end{tabular}
\end{table}

Exemplarily, for step 0 of the algorithm, inspection of the assembly code
for the AVX-512 target shows that a single loop iteration consists of 334
instructions. According to the LLVM machine code analyzer (LLVM-MCA)%
\footnote{\url{https://llvm.org/docs/CommandGuide/llvm-mca.html}},
these are predicted to run with a reciprocal throughput of 248 cycles or an
instruction-per-cycle (IPC) rate of 1.35. According to the analysis, the
main bottleneck is the latency of operations inside the computation of the
power function due to data dependencies. More precisely, the polynomial
evaluation and division operations in the Pad\'e approximation used in the
vectorized \texttt{pow()} implementation~\cite{Fog_2020}, as well as the
extraction of exponents, have long dependency chains.
Since the number of available physical registers and
scheduler windows have limited size to keep around 100-200 instructions in
flight,%
\footnote{%
  The physical register file for floating point numbers in the
  Skylake-X/Cascade Lake architecture has 168 slots, compared to 32
  architectural registers. Similar limits are imposed by the reorder buffer
  (224 entries) and the store buffer (56 entries).
}
little overlap of work from one outer loop iteration (indexed with $i$)
with the next one is possible. Among the 334 instructions, there are 69
fused multiply-add operations, 22 additions/substractions, 31
multiplications, and 3 divisions. Given the LLVM-MCA prediction of
execution in 248 cycles, this corresponds to a throughput of 0.78
arithmetic operations per cycle, or a utilizatoin of 19.6\% of the
arithmetic peak performance. The measured performance of 848 GFlop/s
corresponds to 17\% of the arithmetic peak performance or 85\% of the
predicted arithmetic throughput. This number matches with the ratio of the
measured IPC of 1.32 compared to the predicted IPC of 1.35, showing that
the arithmetic operations have been counted correctly. According to the
roofline model, the memory bandwidth is not a limiting factor for step 0.

Using similar arguments, it can be shown that steps 1, 3, 4, and 5 of
Alg.~\ref{alg:euler} are limited by the in-core execution on the Cascade
Lake processor. Steps 1, 4, and 5 are more strongly effected by long
dependency chains that cannot be overlapped sufficiently with independent
work. This is evidenced by an IPC prediction of 1.27 for the vectorized
\texttt{pow()} function obtained from LLVM-MCA. Step 3 shows a higher
performance that is due to a better instruction-level parallelism obtained
for the evaluation of $\polf(\bUnj)$ and multiplication with $\bc_{ij}$. Step
2 appears odd (see Table~\ref{tab:flop}) due to a high IPC number but
neither high GFlop/s or memory performance. This is because this function
is not vectorized. The alternative of computing all of $d_{ij}$ in
vectorized form via step 1 instead of the symmetrization would be slower
due to the heavy computations in the power function.

\begin{table}
  \caption{Measured run times on a simulation with 29m nodes over 434 time
    steps (1302 RK stage evaluations), run on 96 Intel Skylake cores (2
    compute nodes) at 2.3 GHz with hyperthreading off and on, respectively. Memory
    bandwidth for STREAM triad is 410 GB/s, arithmetic peak 7,066 GFlop/s.}
  \label{tab:supermuc}
  \begin{tabular}{lcccccc}
    \toprule
    & \multicolumn{3}{c}{\textbf{hyperthreading off}} & \multicolumn{3}{c}{\textbf{hyperthreading on}}\\
    \cmidrule(lr){2-4} \cmidrule(lr){5-7}
    & time & arithmetic & bandwidth & time & arithmetic & bandwidth\\
    & [s] & [GFlop/s] & [GB/s] & [s] & [GFlop/s] & [GB/s] \\[0.3em]
    step 0: entropies & 7.49 & 1,094 & 308 & 5.68 & 1,440 & 404 \\
    step 1: offdiagonal $d^L_{ij}$, $\alpha_i$ & 295.4 & 902 & 177 & 210.3 & 1,268 & 257 \\
    step 2: diagonal $d^L_{ii}$, $\tau_n$ & 54.2 & 19 & 352 & 50.8 & 21 & 381\\
    step 3: low-order update & 211.9 & 1,168 & 290 & 212.2 & 1,166 & 307 \\
    step 4: $P_{ij}$, $l_{ij}$ & 250.0 & 726 & 318 & 243.4 & 746 & 338 \\
    step 5: h.-o. update, next $l_{ij}$ & 225.0 & 678 & 317 & 189.9 & 803 & 381 \\
    step 6: final high-order update & 131.3 & 92 & 398 & 135.3 & 89 & 401 \\
    \bottomrule
  \end{tabular}
\end{table}

\subsubsection{Hyperthreading}

In order to further assess the performance bottleneck due to latencies in
the pipelined execution, we run an additional experiment on 96 Intel
Skylake cores comparing enabled and disabled hypertheading; see
Table~\ref{tab:supermuc}. If we run the code with 2-way hyperthreading,
scheduling 96 MPI jobs on each compute node, or 192 jobs in total, performance is
increased for the latency-limited steps of the algorithm. For example, the
run time of step 1 decreases from 295 seconds to 210 seconds, with the
arithmetic throughput reaching 18\% of the arithmetic peak. Similarly,
steps 4 and 5 run considerably faster. On the other hand, step 6 that was
already limited by the memory bandwidth with hyperthreading disabled, is
slightly slower due to additional memory transfer and increased cache
pressure (mainly due to access to transposed entries $l_{ji}$) of the
additional thread running on the same core. The performance with
hyperthreading on the algorithmic step 3 and, to a lesser extent step 4, is
reduced. These steps are affected by additional data streams due to
indirect addressing into the column entries of $\bUnj$ and $\bR^n_j$, which
puts a higher strain on address translation and prefetching.

When comparing the absolute run time of the whole solver (without output)
for 434 time steps of a three-stage Runge--Kutta integrator, we record
1,292 seconds for Skylake without hyperthreading, 1,164 seconds with
hyperthreading, and 1,608 seconds on the slower Intel Cascade Lake system
without hyperthreading. The higher performance of the Intel Skylake system
is in agreement with the hardware specification;
cf.~Table~\ref{tab:systems}. As is expected for an architecture with a
higher machine balance, many of the components run closer to the memory
bandwidth limit. With hyperthreading enabled, step 0, 2, 5 and 6 are now
almost entirely limited by the available memory bandwidth. This shows that
the optimizations presented in this work have paid off.

\subsection{Exploration of algorithmic alternatives}

In order to justify the chosen algorithmic layout, we explore a few
alternative choices and analyze their performance compared to the results
presented in Section~\ref{subse:roofline}.

\subsubsection{Merge step 2 with step 1}

In Alg.~\ref{alg:euler} the symmetry of $d_{ij}^L$ was exploited by only
computing the upper triangular and diagonal portion of $d_{ij}^L$ in step 1
and fixing up the lower triangular part (along with computing the maximal
time-step size) in a separate pass (step 2). The memory access in step 2
is non-contiguous and therefore adds additional memory transfer beyond the
best-case prediction, as can be seen from Table~\ref{tab:memory}.
Given that there is no explicit barrier to fill up the
information, apart from the availability of the upper triangular part, this
step can be done within the loop of step 1. This promises higher performance
because step 1 is limited by the arithmetic operations as described above, so
the additional memory transfer can be expected to be partly hidden. As the
data in Table~\ref{tab:variants_step1} shows, the combined time for steps 1 \&
2 is larger than with doing the transposition as part of the loop. Despite
adding mostly memory access in a core-bound algorithm, there is a small
slowdown compared to step 1 executed alone. This is because the lower
diagonal part $d_{ij}^L$ with $i > j$ for vectorized rows can only be
filled up once the complete upper diagonal part of the matrix has been
computed. Thus, the instruction-level parallelism given an out-of-order
execution window of a few hundreds instructions cannot be fully exploited
while waiting for data that is not already prefetched by the hardware. Even
though this variant provides slightly higher performance, we do not
consider it as the primary algorithm because the basic variant proposed
here only works for an MPI-only parallelization. For parallelization with
threads, the upper diagonal part to read $d_{ji}^L$ is not ready for all
rows, and additional re-ordering or additional computations would be necessary.

\begin{table}
  \caption{Performance comparison of two variants that merge steps 1 and 2
    of Alg.~\ref{alg:euler}: (a) baseline computation with
    Alg.~\ref{alg:euler} as reported in Tables~\ref{tab:memory} and
    \ref{tab:flop}; (b) read transpose values from $d_{ji}^L$ within
    compute loop; (c) compute full row of $d_{ij}^L$ without explointing
    symmetry. All tests were run for 1302 Runge--Kutta stage evaluations on 29
    million grid points with 80 Cascade Lake cores.}
  \label{tab:variants_step1}
  \begin{tabular}{lcccc}
    \toprule
    & time & arithmetic & bandwidth & memory read / write\\
    & [s] & [GFlop/s] & [GB/s] & [doubles / nnz]\\[0.3em]
    \multicolumn{5}{l}{\textbf{(a) Baseline: compute $d_{ij}^L$ as in Alg.~\ref{alg:euler} exploiting symmetry}}\\[0.3em]
    step 1: offdiagonal $d^L_{ij}$, $\alpha_i$ & 391.4 & 681 & 132 & 5.83r + 0.74w \\
    step 2: diagonal $d^L_{ii}$, $\tau_n$ & 62.0 & 17 & 304 & 1.95r + 0.46w \\[0.3em]
    \multicolumn{5}{l}{\textbf{Variant 1: read transpose values from $d_{ji}^L$ within compute loop}} \\[0.3em]
    step 1+2: complete $d_{ij}^L$, $\alpha_i$, $\tau_n$ & 415.7 & 644 & 164 & 7.57r + 1.09w\\[0.3em]
    \multicolumn{5}{l}{\textbf{Variant 2: compute full row of $d_{ij}^L$ without using symmetry}} \\[0.3em]
    step 1+2: complete $d_{ij}^L$, $\alpha_i$, $\tau_n$ & 581.6 & 669 & 164 & 5.45r + 1.09w\\
    \bottomrule
  \end{tabular}
\end{table}

Table~\ref{tab:variants_step1} includes a second variant of the merged
steps 1 and 2 that computes all the entries in $d_{ij}^{L}$ without
considering symmetry. While the data access is lowest in this case with
loads that are mostly streaming, the performance is significantly lower due
to the increased number of computations.

Note that writing into $d_{ij}^{L}$ can be done with streaming stores for
the baseline algorithm as well as variant 2, where the full $d_{ij}^L$
matrix is computed, whereas regular stores with 1 double with
read-for-ownership transfer is needed for variant 1 to be able to hit parts
of the transposed access in cache.

\subsubsection{Split computation of $P_{ij}$ into steps 3 and 4}

The contribution $(d_{ij}^{H,n}-d_{ij}^{L,n})(\bUnj-\bUni)$ to matrix $P_{ij}$
is already available in step 3 of the algorithm, whereas the baseline
algorithm recomputes this information in step 4. Given that both step 3 and 4
are limited by the computations in the core, an algorithmic alternative is to
store this temporary result in the storage location of $P_{ij}$ in step 3 and
re-load it for the computation of step 4. This incurs writes of five
doubles in step 3 (which can be done with streaming stores) and reads of up
to four doubles in step 4. On the other hand, $d_{ij}^{L,n}$ and $\bUnj$ do
not need to be loaded again in step 4. This modification is reported as
variant 3 in Table~\ref{tab:variants_p_ij} (b). The results clearly show
the additional write data transfer in step 3 and the read transfer in step
4, with both steps running more slowly.
The computational time of both steps is significantly increased and the
steps are now mostly memory transfer limited. The measured throughput of
around 300 GB/s is slightly below the STREAM triad limit of the platform.

While this algorithmic variant is not profitable on the chosen hardware, it
can be promising for hardware with high bandwidth-memory interfaces, or when
indirect addressing (for example, access to $\bUnj$) is more expensive.

\begin{table}
  \caption{Performance comparison of different variants for computing
    $l_{ij}$ and $P_{ij}$: (a) baseline computation with
    Alg.~\ref{alg:euler} as reported in Tables~\ref{tab:memory} and
    \ref{tab:flop}; (b) split computation of $P_{ij}$ into steps 3 and 4 in
    order to to reduce indirect addressing and computations; (c) compute
    both $l_{ij}$ and $l_{ji}$ rather than symmetrizing over the memory
    access the computation of $P_{ij}$ into steps 3 and 4 in order to
    reduce indirect addressing and computations; (d) do not store the
    matrix $P_{ij}$ and instead compute the entries on the fly from the
    respective ingredients in steps 5 and 6 of Alg.~\ref{alg:euler}. All tests
    were run for 1302 Runge--Kutta stage evaluations on 29 million grid points with
    80 Cascade Lake cores and report measured data with LIKWID.}
  \label{tab:variants_p_ij}
  \begin{tabular}{lcccc}
    \toprule
    & time & arithmetic & bandwidth & memory read / write\\
    & [s] & [GFlop/s] & [GB/s] & [doubles / nnz]\\[0.3em]
    \multicolumn{5}{l}{\textbf{(a) Baseline: compute $l_{ij}$ and  $P_{ij}$
    as in Alg.~\ref{alg:euler}}}\\[0.3em]
    step 3: low-order update & 277.0 & 892 & 222 & 7.24r + 0.53w\\
    step 4: $P_{ij}$, $l_{ij}$ & 317.8 & 571 & 248 & 4.43r + 6.03w\\
    step 5: h.-o. update, next $l_{ij}$ & 268.7 & 568 & 260 & 8.06r + 1.20w\\
    step 6: final high-order update & 132.5 & 91 & 394 & 6.71r + 0.21w\\[0.3em]
    \multicolumn{5}{l}{\textbf{(b) Variant 3: split $P_{ij}$ into two parts}} \\[0.3em]
    step 3: low-order update, first half of $P_{ij}$ & 325.2 & 779 & 312 & 7.33r + 5.52w\\
    step 4: second half of $P_{ij}$, $l_{ij}$ & 358.4 & 468 & 292 & 7.75r + 6.00w\\[0.3em]
    \multicolumn{5}{l}{\textbf{(c) Variant 4: compute both $l_{ij}$ and $l_{ji}$}} \\[0.3em]
    step 4: $P_{ij}$, $\min(l_{ij},l_{ji})$ & 558.7 & 567 & 152 & 5.78r + 6.18w\\
    step 5: h.-o. update, next $l_{ij}$ & 259.2 & 584 & 204 & 5.74r + 1.09w\\[0.3em]
    \multicolumn{5}{l}{\textbf{(d) Variant 5: compute $P_{ij}$ on the fly}} \\[0.3em]
    step 4: $l_{ij}$ & 272.2 & 669 & 144 & 4.33r + 1.07w\\
    step 5: h.-o. update, next $l_{ij}$ & 389.1 & 507 & 165 & 7.14r + 1.21w\\
    step 6: final high-order update & 210.9 & 271 & 239 & 6.30r + 0.23w\\
    \bottomrule
  \end{tabular}
\end{table}

\subsubsection{Compute symmetrization of limiter matrix}

In steps 5 and 6, the update of $P_{ij}^n$ requires the operation
$\min(l_{ij},l_{ji})$, with the latter accessing transpose entries in the
matrix. In order to reduce the memory transfer, we analyze a variant 4 of our
baseline algorithm that adds the computation of $l_{ji}$ within step 4 of the
algorithm. Given that the matrix $P_{ij}$ is skew-symmetric in the sense
$\lambda m_i P_{ij} = - \lambda m_j P_{ji}$, only an additional load to $\bUnj$
and $\texttt{bounds}_j$ is needed, in addition to the actual computation in
\texttt{limiter.compute}. Table~\ref{tab:variants_p_ij} (c) shows an
implementation of this variant. While the run time of step 5 and the
associated memory access are slightly reduced because the transposed
entries are not needed, we observe a noticable increase in execution time
in step 4 becausethe simultaneous computation of $l_{ij}$ and $l_{ji}$ in
step 4 doubles the number of critical computations. As discussed
previously, latency effects inside the limiter are the dominant bottleneck,
which explains why the additional computations do not increase the
arithmetic throughput. Overall, this option is less attractive because the
time gained in step 5 is only minor, given that the gain is mostly due to a
reduction in stalls when waiting for the indirectly accessed column data
$l_{ji}$ to arrive. A similar modification could be considered for
computing the next $l_{ij}$ and $l_{ji}$ in anticipation of step 6. This
has similar deficiencies as the alternative discussed above, and in
addition needs to wait for the update $\bU^{n+1}_i$ to be finished for all
columns $\mathcal I(j)$.

\subsubsection{Computation of entries of $P_{ij}$ on the fly}

As a final algorithm variant 5, we consider to skip the storage of $P_{ij}$
and instead evaluated it by the formula
$\bP_{ij} = \frac{\tau_n}{\lambda
m_i}\Big(\big(d_{ij}^{H,n}-d_{ij}^{L,n}\big)\big(\bUnj -
\bUni\big)+b_{ij}\bR_j-b_{ji}\bR_i\Big)$
whenever necessary. This significantly reduces the memory access as the
matrix $P_{ij}$ amounts to a read/write of five doubles per non-zero entry,
compared to the two matrices $d_{ij}^{L,n}$ and $m_{ij}$ (for computing
$b_{ij}$) and the vectors $\bUni$, $\bUnj$ as well as $\bR_i^{n}$,
$\bR_j^{n}$. Table~\ref{tab:variants_p_ij} (d) compares this variant with
the baseline algorithm. While step 4 becomes 45 seconds faster by removing
the expensive write operation of $P_{ij}$, the additional computations slow
down steps 5 and 6 by 120 seconds and 78 seconds, respectively. From the
recorded memory transfer, it becomes clear that the gain in transfer is not
too high, which can be explained by the fact that besides the two matrices
$d_{ij}^{L,n}$ and $m_{ij}$ also indirect addressing to $\bUnj$ and
$\bR^n_j$ needs to be performed. As discussed previously, additional data
that is kept in flight increases pressure on the caches and also cache misses,
eliminating part of the gain.

\subsection{Strong scaling}

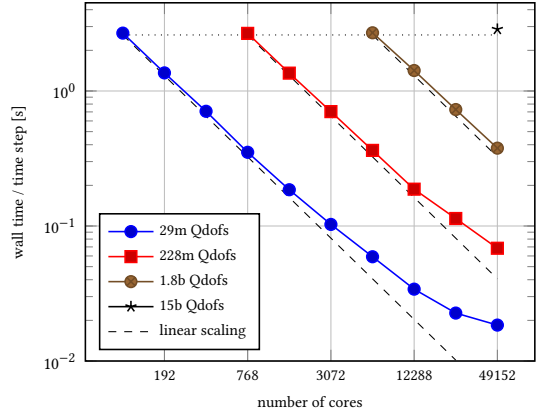
\begin{figure}
  \begin{tikzpicture}
    \begin{loglogaxis}[
      width=0.5\textwidth,
      height=0.42\textwidth,
      xlabel={number of cores},
      ylabel={wall time / time step [s]},
      tick label style={font=\scriptsize},
      label style={font=\scriptsize},
      legend style={font=\scriptsize},
      legend cell align=left,
      legend pos={south west},
      grid,
      mark size=2.2,
      xtick={192,768,3072,12288,49152},
      xticklabels={192,768,3072,12288,49152},
      ymin=1e-2,ymax=4.5e+0,
      semithick
      ]
      \addplot table[x expr={\thisrowno{0}*48}, y expr={\thisrowno{3}/\thisrowno{2}}] {\tableStrongA};
      \addlegendentry{29m Qdofs};
      \addplot table[x expr={\thisrowno{0}*48}, y expr={\thisrowno{3}/\thisrowno{2}}] {\tableStrongB};
      \addlegendentry{228m Qdofs};
      \addplot table[x expr={\thisrowno{0}*48}, y expr={\thisrowno{3}/\thisrowno{2}}] {\tableStrongC};
      \addlegendentry{1.8b Qdofs};
      \addplot table[x expr={\thisrowno{0}*48}, y expr={\thisrowno{3}/\thisrowno{2}}] {\tableStrongD};
      \addlegendentry{15b Qdofs};
      \addplot[dashed,black,thin] coordinates {
          (96,2.6)
          (24576,2.6/256)
        };
        \addlegendentry{linear scaling};
      \addplot[dashed,black,thin] coordinates {
          (768,2.6)
          (49152,2.6/64)
        };
      \addplot[dashed,black,thin] coordinates {
          (6144,2.6)
          (49152,2.6/8)
        };
      \addplot[dotted,black,thin] coordinates {
          (96,2.6)
          (49152,2.6)
        };
      \end{loglogaxis}
    \end{tikzpicture}
  \caption{Strong and weak scaling of solver on Intel Skylake for problem
  sizes between 29 million and 15 billion points. The data has been gather by
  runs using between 434 and 1732 time steps using a three-stage Runge--Kutta
  scheme and report the run time per time step.}
  \label{fig:strong}
\end{figure}

Since the solver only involves local communication to the neighbors via
non-blocking MPI send commands, plus one \texttt{MPI\_Allreduce} for computing
the time step size, it is straight-forward to run the solver for
simulations on large supercomputers. Fig.~\ref{fig:strong} shows the result
of a strong scaling experiment on up to 1,024 compute nodes of Intel Skylake on the
SuperMUC-NG machine in Garching, Germany. The experiment is conducted with
2-way hyperthreading enabled using a separate MPI rank for each core and
two threads per core. The largest computations are run on 49,152 MPI ranks
with 98,304 threads in total. The times reported in this section are based
on the minimal time recorded for four runs of the complete time evolution
to minimize disturbances from other jobs running on the machine.

\pgfplotstableread{
  nodes  dofs     nts time
  2      28656640 434 1164.22
  4      28656640 434 590.74
  8      28656640 434 306.99
  16     28656640 434 152.53
  32     28656640 434 80.57
  64     28656640 434 44.58
  128    28656640 434 25.69
  256    28656640 434 14.78
  512    28656640 434 9.83
  1024   28656640 434 8.01
}\tableStrongA
\pgfplotstableread{
  nodes  dofs     nts time
  16     227870720 660 1762.82
  32     227870720 867 1175.42
  64     227870720 867 609.91
  128    227870720 867 314.82
  256    227870720 867 162.47
  512    227870720 867 98.54
  1024   227870720 867 59.33
}\tableStrongB
\pgfplotstableread{
  nodes  dofs     nts time
  128    1817448448 650 1753.9
  256    1817448448 1250 1772.13
  512    1817448448 1732 1262.17
  1024   1817448448 400 150.89
}\tableStrongC
\pgfplotstableread{
  nodes  dofs     nts time
  1024   14517542912 500 1432.55
}\tableStrongD

The results in Fig.~\ref{fig:strong} show an almost perfect scaling to
times of around 0.2 seconds per time step or 0.07 seconds per Runge--Kutta
stage. The smallest size with 28 million nodes continues to improve
throughput all the way to 49k cores with 0.018 seconds per time step.
However, the parallel efficiency drops to 46\% already for 24k cores, using
the run with 1.8 billion unknowns on the same core count as baseline. If we
define the strong scaling limit as the point where 80\% of the saturated
performance is obtained \cite{Fischer_2020}, the 29m grid point case scales
to 3072 cores (with 81\% of parallel efficiency) and the 228m grid point
case scales to 12k cores with 89\% parallel efficiency.  This excellent
scalability is the result of judicious algorithmic choices with the
majority of communication only between nearest neighbors in the mesh. In
each Runge--Kutta stage, one \texttt{MPI\_Allreduce} operation is also
necessary to control the time step size.

The lowest computational time per Runge--Kutta stage is around $5\times
10^{-3}$ seconds for the proposed algorithm. We can compare this number
with the time for one CG iteration of a matrix-free solvers of $2\times
10^{-4}$ seconds on the same SuperMUC-NG system \cite[Fig.~8]{Arndt_2020}
for the benchmark described in \cite{Fischer_2020} or $10^{-4}$ seconds for
the nearest-neighbor communication of a matrix-vector product
\cite{Kronbichler_2018}. The higher limit for scaling in our case can be
explained by the significantly more expensive stencil update, as each
update involves seven nearest-neighbor communication steps for the various
intermediate quantities in the algorithm and one global reduction, which
already explains a factor of around ten in the time increase. Furthermore,
the computation on 29 million mesh points on 49k cores corresponds to only
290 mesh points per thread, which in itself is a very low value for any
PDE-parallel code. Thus, the task granularity is very small at this point,
which makes small imbalances in the SIMD/non-SIMD portions more difficult
to control. Also, latency effects in the various algorithmic stages,
including warm-up of the instruction caching, also play a role at this
level. We leave possible improvements along the strong scaling limit to
future work (see Remark~\ref{rem:overlap}).


\section{Conclusion and outlook}
\label{sec:conclusion}

In this paper we have discussed the efficient implementation of a
second-order collocation-type finite-element compressible Euler solver. To
this end we started with the \emph{mathematical description} of the scheme
that is guaranteed stable without the need of any tuning parameters. We
then reorganized and optimized the given algorithmic structure
(Sec.~\ref{sec:scheme}) and discussed a scalable high-performance
implementation (Sec.~\ref{sec:implementation}).
The main algorithmic building blocks are traversals through CELL-based sparse
matrices with indirect addressing into the solution vector and some
auxiliary quantities, as well as a relatively high density of division and
transcendental power functions.
We demonstrated excellent
arithmetic throughput and scaling (Sec.~\ref{sec:benchmarks}) and justified
our algorithmic choices against alternatives. We point out a
number of possible optimizations that we have not pursued and
that we leave for further research and development:
\vspace{-0.5em}
\begin{itemize}
  \item
    Further reduction of the number of MPI synchronizations by increasing
    the overlap of shared cells between neighboring MPI ranks. In our
    current implementation the overlap is one ghost layer of cells
    \cite{dealiicanonical}. An increased overlap would allow to remove
    most of the synchronization steps outlined in
    Fig.~\ref{fig:mpi-synchronization}.
  \item
    More efficient coefficient computation of transcendental functions by
    using a better pipelined custom vectorized \texttt{pow()}
    implementation as discussed in Sec.~\ref{subse:vectorized_pow}.
  \item
    The developed algorithmic structure and the use of a SELL-based sparse
    matrix format for storage gives hope that the proposed algorithms
    will also perform reasonably on GPU systems or other HPC architectures.
    Performance-portable implementations, such as realizations with Kokkos
    \cite{Kokkos_2014} or Raja \cite{Raja_2014}, for this kind of equations
    are still missing, but could be guided by the performance envelopes and
    algorithmic behavior identified in the present contribution.
\end{itemize}
By allowing to modify the \emph{mathematical structure} we expect an even larger
gain in performance of the algorithm:
\vspace{-0.5em}
\begin{itemize}
  \item
    The 3D stencil for lowest-order $\mathcal{Q}_1$ elements has 27
    entries. It is an open research question whether it is possible to
    reduce the stencil size (for example by
    additional lumping) for part of the for loops in Alg.~\ref{alg:euler}.
    In addition, the convex-limiting methodology
    \cite{gnpt_second_order_2018} is not restricted to a CG discretization and
    can be also applied to (high-order) DG discretizations
    \cite{Zalesak_2005,Pazner_2020}. Such flux-corrected DG schemes might
    promise a higher arithmetic throughput and more regular data access.
  \item
    Much of the computational bottleneck stems from the heavy use of the
    transcendental \texttt{pow()} function. An investigation of modified
    limiter approaches that still guarantee the invariant-domain property
    but use a cheaper to evaluate 3-convex function $\Psi(\bU)$
    (Sec.~\ref{subsec:specific-entropy-limiter}) thus seems very tempting.
  \item
    A similar consideration can be made for the entropy-viscosity
    commutator (Sec.~\ref{subsec:entropy-viscosity-commutator}) and the
    subsequent interaction with the limiter: it needs to be investigated
    whether the number of
    transcendental functions in the indicator can be reduced by potentially
    including certain entropies in the limiting process or by using
    monotonicity/convexity in some functional relations to pull out power functions
    from the inner $j$ loop to the outer $i$ loop.
\end{itemize}


\begin{acks}
  MK was supported by the Bayerisches Kompetenznetzwerk f\"ur
  Technisch-Wissenschaftliches Hoch- und H\"ochstleistungsrechnen
  (KONWIHR). MM acknowledges partial support by the NSF under grant
  DMS-1912847. The authors gratefully acknowledge the Gauss Centre for
  Supercomputing e.V.~(\texttt{www.gauss-centre.eu}) for funding this
  project by providing computing time on the GCS Supercomputer SuperMUC at
  Leibniz Supercomputing Centre (LRZ, \texttt{www.lrz.de}) through project
  id pr83te. This material is based upon work supported by a
  ``Computational R\&D in Support of Stockpile Stewardship'' grant from
  Lawrence Livermore National Laboratory, the Air Force Office of
  Scientific Research, USAF, under contract number FA9550-15-1-0257.
\end{acks}


\end{document}